\documentclass[12pt,fleqn,psfig]{article}
\usepackage{amsmath, amssymb, graphicx}
\usepackage[hang,scriptsize,bf]{caption}
\usepackage{cite}

\usepackage{geometry}
\usepackage{color}
\usepackage[latin1]{inputenc}
\usepackage{mathrsfs}
\usepackage{amsfonts}
\usepackage[title]{appendix}

\begin{document}

 
\thispagestyle{empty}
\renewcommand{\thefootnote}{\fnsymbol{footnote}}

\begin{center} \noindent {\bf  Geometry and Perturbative Sensitivity of\\ non-Smooth Caustics of the Helmholtz Equation}\footnote{DARPA Distribution Statement ``A'' (Approved for Public Release, Distribution Unlimited.)}

\end{center}

\bigskip\bigskip\bigskip

\bigskip\bigskip\bigskip

\begin{center}
Z.~Guralnik\footnote{Zachary.S.Guralnik@leidos.com}, C.~Spofford\footnote{Charles.W.Spofford@leidos.com} and K.~Woolfe\footnote{Katherine.F.Woolfe@leidos.com}\\
\it \small Leidos, Inc. 11951 Freedom Dr, Reston, VA, 20190
\end{center}
\bigskip

\bigskip\bigskip\bigskip\bigskip

\renewcommand{\thefootnote}{\arabic{footnote}}

\centerline{ \small Abstract}
\bigskip

\small 

The geometry of non-smooth $A_{n>2}$ caustics in solutions of the Helmholtz equation is analyzed using a Fock-Schwinger proper-time formulation.
In this description, $A_3$ or cusp caustics are intimately related to poles of a quantity called the einbein action in the complex proper-time, or einbein, plane.  The residues of the poles vanish on spatial curves known as ghost sources, to which cusps are bound. The positions of cusps along the ghost sources is related to the value of the poles.  A similar map is proposed to relate essential singularities of the einbein action to higher order caustics.  The singularities are shown to originate from degenerations of a certain Dirichlet problem as the einbein is varied.  It follows that the singularities of the einbein action, along with the associated aspects of caustic geometry, are invariant with respect to large classes of perturbations of the index of refraction.


\newpage
\section{Introduction}

Regions of intense fields in the neighborhood of caustics are a generic feature of wave propagation problems.  There are a multitude of disparate contexts in which caustics are of interest, including   acoustics, optics,  and gravitational lensing to name just a few. 
In this article we analyze the sensitivity of non-smooth caustics to perturbations of the index of refraction in the Helmholtz equation, in the course of which we further develop a description proposed in \cite{GLefschetz} relating non-smooth caustics to singularities of a quantity referred to as the `einbein action'.   Since caustics arise due to the focal behavior of classes of rays sweeping large spatial regions away from the caustic, one might suspect an inherent stability against local or oscillatory perturbations of the index of refraction.  In a qualitative sense, the size of this region is larger for higher order caustics than for a point on a smooth $A_2$, or `fold', caustic.  This stability is born out by quantitative arguments based on the effect of perturbations of the index of refraction on singularities of the einbein action.

   
The einbein action formulation of solutions of the Helmholtz equation is essentially a Schwinger-Fock-Feynman proper-time solution \cite{Fock,Schwinger,Feynman1,Feynman2,Palmer,Schlottmann}, in which the integration over positive real proper time is generalized to include paths in the complex plane \cite{GLefschetz}. This approach will be reviewed briefly in section \ref{einbeinreview}. The wave field can be written as,
\begin{align}
\phi(\vec x) &= \int_{\Gamma} d\Lambda \Psi(\vec x,\Lambda)\\
&\Psi \equiv e^{i\mathbb S}
\end{align}
where $\mathbb S$ is the einbein action and $\Lambda$ is the Schwinger proper time, also referred to as the einbein.  The origin of the einbein nomenclature is a relation between the eikonal equation of ray theory and reparameterization invariance enforced by a Lagrange multiplier, the einbein, in a path integral solution of the Helmholtz equation.  Gauge fixing and reversing an order of integration in the path integral, the einbein becomes the Schwinger proper time.  We shall generally refer to the Schwinger proper time as the einbein, for the sake of brevity and in keeping with \cite{GLefschetz}.
Although the integration contour $\Gamma$ is ordinarily the positive real axis, it is more generally taken to be a sum over steepest descent complex paths connecting essential singularities of $\Psi(\Lambda)$, including that at infinity.  In the soluble examples described in \cite{GLefschetz}, the essential singularities of $\Psi(\Lambda)$ are simple poles of $\mathbb S(\Lambda)$, however $\mathbb S(\Lambda)$ may contains essential singularities as will be seen in examples considered here.  The particular choice of segments making up the integration contour depends on  the boundary conditions.  There is a map between each of the steepest descent paths and eigenrays, including complex eigenrays such as those described in \cite{Keller,Chapman,KravtsovOrlov1}.  The analytic structure of the field $\phi$ as a function of the parameters of the problem was shown \cite{GLefschetz} to be related to monodromies mapping inequivalent classes of integration contours into each other as these parameters are varied.  The singularities of the einbein action in the complex $\Lambda$ plane play a crucial role in these monodromies, and also in the description of higher order caustics described in detail here.    

The field at a caustic, including the surrounding neighborhood in which geometric optics is valid,  is given by a uniform asymptotic approximation \cite{ Maslov,Kravtsov,Duistermaat,Berry2,
KravtsovOrlov2,KravtsovOrlov1,Pearcey,
Chester,Kravtsov1,Kravtsov2,Ludwig,Berry4,
Berry,Berry3}.   In this approximation, $\phi(\vec x) $ is represented as a sum  of a function $f(a_I)$ and its derivatives, where
\begin{align}
f(a_I) =\int d\lambda e^{iP(\lambda,a_I(\vec x))}\, 
\end{align}
and $P$ is a polynomial in $\lambda$ known as the catastrophe generating function, with coefficients  $a_I$ depending on $\vec x$. 
The generating function $P(\lambda)$ is related to the einbein action $\mathbb S(\Lambda)$ by a map $\lambda(\Lambda)$ which is valid for $\vec x$ in the neighborhood of a caustic.  The einbein action can be thought of as a generalization of the generating function which is valid globally, including multiple caustics.  A polynomial in a single variable suffices to describe caustics of corank one, or the $A_n$ type.  These include the fold caustic $A_2$,  the cusp caustic  $A_3$, the swallowtail caustic and $A_4$ and the butterfly caustic $A_5$.  An einbein action description of caustics with corank greater than one, such as the $D_N$ and $E_N$ caustics which involve a polynomial in two variables $\lambda_{1,2}$, is currently unknown. 


In \cite{GLefschetz}, it was conjectured that cusp caustics lie at points $\vec x$ within curves on which the residues of the poles of $\mathbb S(\Lambda)$ vanish.  These curves were dubbed ghost sources, for reasons to be reviewed here.  Equivalently, one can say that cusp caustics lie at the point at which fold caustics meet ghost sources.  While the intersection of a ghost source with a fold caustic is accompanied by three coalescing critical points,  there is currently no proof that all cusp caustics arise at such intersections.  We shall find evidence that the location of cusps along ghost source curves is determined in part by the location of the associated pole, referred to as a ghost pole, in the complex $\Lambda$ plane. 
The exactly soluble examples, such as those discussed in \cite{GLefschetz}, do not include  swallowtail and butterfly caustics.  We shall consider perturbations of the exactly soluble cases hinting that these higher order caustics are related to essential singularities of the einbein action rather than poles.  

The einbein action can be obtained from a path integral 
\begin{align}
e^{i\mathbb S} = \int {\cal D} \vec X(\tau) e^{i\int_0^1 d\tau {\cal L}\left(\Lambda,\vec X(\tau),\frac{d\vec X}{d\tau}\right)}\, ,
\end{align}
where the fixed endpoints $\vec X(0),\vec X(1)$ are the arguments of the Green's function of the Helmholtz equation. The origin of the singularities of $\mathbb S(\Lambda)$ can be traced to singularities of the Dirichlet problem for the Euler Lagrange equations derived from ${\cal L}$. Although the solutions of these equations are spatial trajectories,  they depend on $\Lambda$ and are not equivalent to rays. 
Approaching special values of $\Lambda$ which correspond to singularities of $\mathbb S$, solutions of the Dirichlet problem for generic endpoints diverge, whereas  continuous classes of finite solutions solve the Dirichlet problem for special endpoints corresponding to the ghost sources. 
The divergence of solutions of the Dirichlet problem for generic boundary conditions has important implications for the effect of perturbations of the index of refraction.  In fact for large classes perturbations localized to spatial regions which the diverging solutions evade, there is no effect on the singularities of the einbein action or the corresponding aspects of caustic geometry. The principle effect of such perturbations is only to slide caustics along ghost sources.  

If the only singularities of the einbein action were poles,  as seems to be the case in the exactly soluble instances,  the singularities would be essentially immutable.  For reasons to be discussed, no local perturbation of the index of refraction can continuously move a pole or change its residue.
However it will be shown that essential singularities of the einbein action arise naturally under perturbation, and that these may vary under local perturbations. Numerical evidence also suggests that these essential singularities are related to higher order caustics $A_{n>3}$.  Remarkably, the pattern of degeneration for the Dirichlet problem in the space spanned by $\Lambda$ and the boundary conditions very closely resembles the spatial pattern of the corresponding caustic.  This can be seen numerically, although a mathematical proof of the apparently simple map between these two spaces is currently lacking.  

The organization of this article is as follows. Section \ref{einbeinreview} contains a review of the einbein, or Schwinger proper time, formalism discussed in detail in \cite{GLefschetz}. The relation between poles in the complex einbein plane and the geometry of cusp caustics will be reviewed in section \ref{simplecusp}, in the context of a very simple example for which the einbein action is exactly computable and has a single ghost pole. 
An exactly soluble model having an infinite number of ghost poles is described in section \ref{polepairs}.  In this case it is shown that each cusp is associated with a particular ghost pole. Both the cusp locations and the geometry of the attached fold caustics in the immediate neighboorhood of the cusp are accurately captured by  effective einbein actions neglecting all but one ghost pole.  This reduced construction is analogous to the uniform asymptotic description in terms of the Pearcey function, although it is formulated directly in terms of spatial coordinates rather than the arguments $\zeta_I$ of the Pearcey function, which are coefficients in the generating polynomial for the cusp caustic. The map $\zeta_I(\vec x)$ is constructed in section \ref{Pearceymap}. 
Section \ref{origin} describes the origin of poles and ghost sources in the einbein action in terms of singularities of a Dirichlet problem.  The degeneration of the Dirichlet problem is shown to predict the location of cusp caustics for a case which is not exactly soluble, namely the the index of refraction known as the Munk profile, of common use in ocean acoustics. The effect of perturbations of the index of refraction on singularities of the einbein action is considered in section \ref{perttheory}, along with the map between degenerations of the Dirichlet problem,  essential singularities, and  $A_{n>3}$ caustics.
Finally, the insensitivity of ghost sources to the index of refraction in a Laurent series expansion of the einbein action is shown, neglecting the possibility of essential singularities.

\section{The einbein formulation}\label{einbeinreview}

The einbein formulation of the solution of the Helmholtz equation is described in detail in \cite{GLefschetz}.  Here we review some of the key features relevant to the geometry of cusp caustics.
The Helmholtz equation
\begin{align}\label{Helmholtz}
\left(\vec\nabla_x^2 + k_0^2 n(\bf x)^2\right)\phi(\vec x) = J(\vec x) 
\end{align}
can be solved starting from the solution of a Shr\"oedinger equation 
 \begin{align}\label{Schrod}
\frac{i}{k_0} \frac{\partial}{\partial\Lambda}\Psi+ \left(\frac{1}{k_0^2}\vec{\nabla}_x^2 +  n(\vec x)^2\right)\Psi =0\, ,
\end{align}
since \eqref{Schrod} implies
\begin{align}\label{surff}
\left(\vec{\nabla}_x^2 +  k_0^2 n(\vec x)^2\right) \int_\Gamma d\Lambda \Psi =  -ik_0 \Psi \big |_{\partial\Gamma}\, .
\end{align}
One can then write 
\begin{align}
\phi(\vec x) = \int_{\Gamma} d\Lambda\, \Psi(\Lambda,\vec x)\, ,
\end{align}
for any complex contour $\Gamma$ such that
\begin{align}
J = \left.\Psi\right|_{\partial \Gamma}\, .
\end{align}
This is essentially the Schwinger proper time solution, except that in the  conventional approach the integration is restricted to the positive real $\Lambda$ axis,  yielding the Green's function with radiation boundary conditions.  The full solution set is more general, involving sums over in-equivalent complex contours.  For the source-free Helmholtz equation, the contour $\Gamma$ must either enclose singularities in the complex $\Lambda$ plane, or have endpoints at which $\Psi$ vanishes.  For a delta function source,  $\Gamma$ must include an endpoint at which $\Psi$ vanishes for all $\vec x$ except the location of the source.  The standard integration over positive real $\Lambda$ is just marginally convergent, but is equivalent to an integral over  sums of certain complex steepest descent paths connecting essential singularities of $\Psi$, approached such that $\Psi$ vanishes. In the exactly soluble cases described in \cite{GLefschetz}, the essential singularities of $\Psi$ are poles of the einbein action $\mathbb S$ where 
\begin{align}
\mathbb S \equiv -i\ln\Psi\, .
\end{align}
As  $\vec x$ approaches the location of a source, the residue of a pole of $\mathbb S$ at $\Lambda=0$ vanishes.

To illustrate the analytic structure of $\Psi$ for a simple example described in \cite{GLefschetz}, consider the Helmholtz equation in two dimensions with index of refraction and source given by 
\begin{align}
n(X,Z)^2 &= n_0^2 -aZ\\
J(X,Z) &= \delta(X)\delta(Z) \label{detl}\, .
\end{align}
In this case,
\begin{align}\label{einbone}
\mathbb S =  ik_0\left(\frac{1}{4\Lambda}(x^2 + z^2) + \Lambda (n_0^2- \frac{a}{2} Z) - \frac{1}{12}a^2\Lambda^3\right) + {\rm logarithmic\,\, terms} \, .
\end{align}
The logarithmic terms, which are ${\cal O}(1)$ in the large $k_0$ limit, are generally irrelevant to the questions of interest here and will be frequently ignored. In this example, 
there is third order pole at $\Lambda=\infty$ and a first order pole at $\Lambda=0$, which may serve as endpoints of integration. The pole at infinity must be approached within the three angular domains satisfying $\pi<arg(\Lambda^3)<2\pi$ for positive real $a$, while the pole at $\Lambda=0$ must be approached within the angular domain
$\pi<arg(\Lambda)<2\pi$.  Note that the residue of the $\Lambda=0$ pole vanishes at $x=z=0$, the location of the source.  At the point $\vec x=0$, the essential singularity of $\Psi$ at $\Lambda=0$ is absent, such that $\Psi$ no longer vanishes as $\Lambda\rightarrow 0$.  For a contour ending at this pole, there is a boundary contribution in \eqref{surff} which can be shown to equivalent to the delta function source \eqref{detl}. 
In fact for any index of refraction and source $J=\delta(\vec x- \vec x')$, the einbein action invariably has a pole of the form
\begin{align}
\mathbb S = \frac{( \vec x - \vec x')^2}{4\Lambda} + \cdots
\end{align}
with the remaining terms in the action dependent on the index of refraction $n(\vec x)$.  

\begin{figure}[!h]
	\center{
		\includegraphics[width= 200pt]{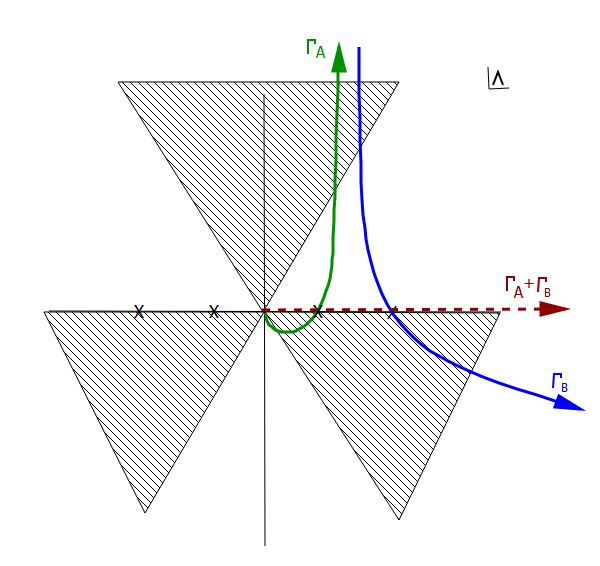}
		\caption{Steepest descent  contours $\Gamma_A$ and $\Gamma_B$ in the complex $\Lambda$ plane, for $\vec x$ in the illuminated zone.  These contours are bounded by poles of the einbein action, and pass through real critical points. The critical point associated with $\Gamma_A$ maps to the eigenray $A$  of figure \ref{fig:TwoRays} which has not touched the caustic, whereas the critical point along $\Gamma_B$ maps to the eigenray $B$ which intersects the caustic. The Greens function is a sum of integrals of $\Psi$ along these contours, equivalent to an integral along the positive real axis. The shaded wedges denote the angular domains at infinity within which integrals of $\Psi$ converge.}
		\label{fig:Contours1}
	}
\end{figure}

\begin{figure}[!h]
	\center{
		\includegraphics[width= 200pt]{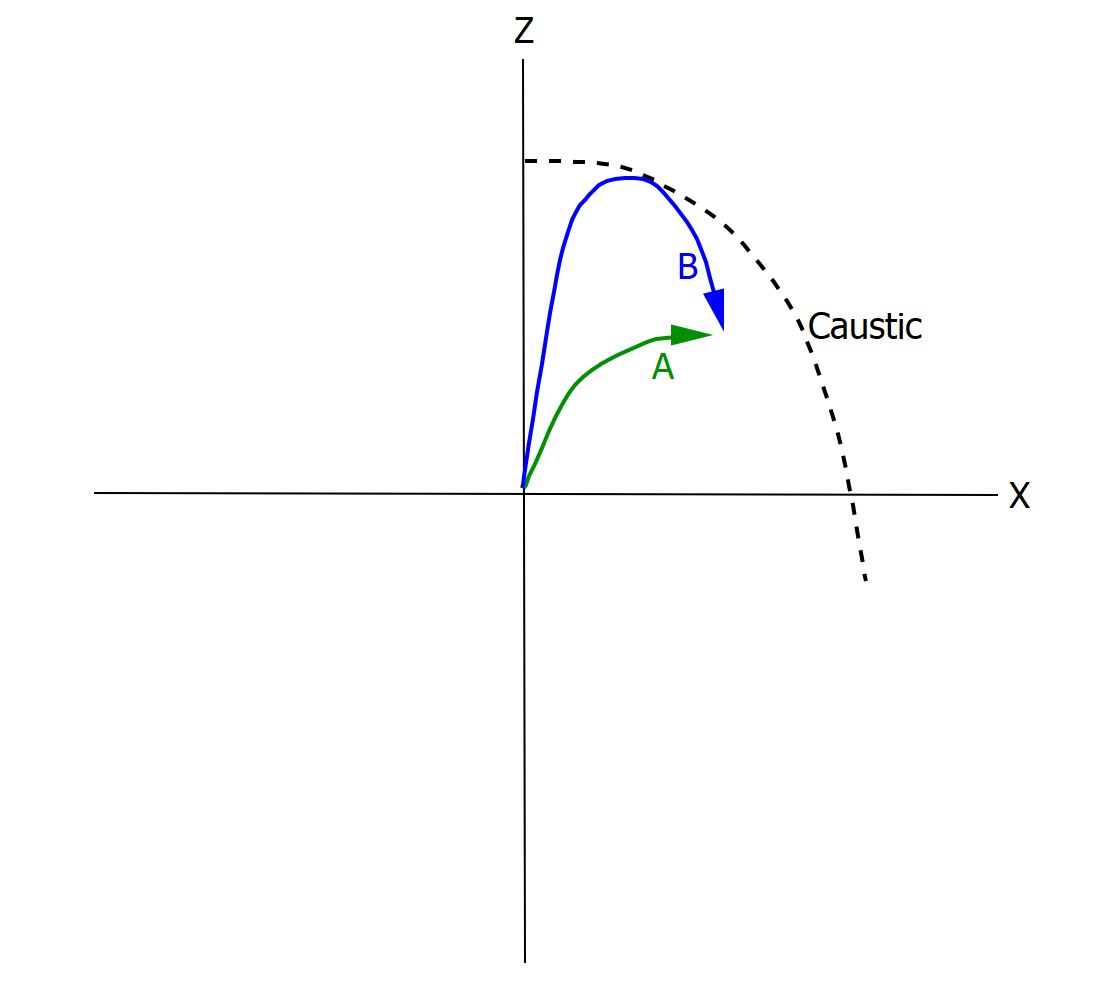}
		\caption{Two eigenrays ending at the same point $\vec x$ in the illuminated zone, for the index of refraction $n(x,z)^2=n_0^2-Az$ and a source at the origin.}
		\label{fig:TwoRays}
	}
\end{figure}

\begin{figure}[!h]
	\center{
		\includegraphics[width= 200pt]{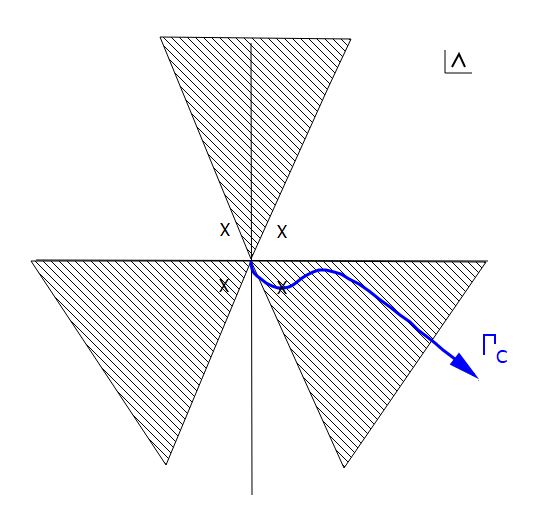}
		\caption{A steepest descent integration contour $\Gamma_C$ in the complex $\Lambda$ plane for $\vec x$ in the shadow zone, connecting the pole of the einbein action at $\Lambda=0$ to that at infinity, and equivalent to the marginally convergent integration contour over the positive real axis. The contour passes through a complex critical point, marked with  'X' associated with a complex eigenray. The shaded wedges denote the angular domains within which contours extending to infinity yield convergent integrals.}
		\label{fig:Contours2}
	}
\end{figure}

Steepest descent, or constant ${\rm Re}(\mathbb S)$, contours associated with \eqref{einbone} are shown in figure \ref{fig:Contours1}, for $\vec x$ within the `illuminated zone' in which there are two real eigenrays. The full contour is a sum of two segments,  one of which corresponds to the eigenray which has touched the fold caustic along the curve,
\begin{align}
 n_0^4 -a z n_0^2 - \frac{1}{4}a^2 x^2  =0\, ,
\end{align}
while the other contour corresponds to the ray which has not touched the caustic, as shown in
in figure \ref{fig:TwoRays}.
In the shadow zone, where are there are no real eigenrays, the steepest descent contour has one component, passing through a complex critical point, as shown in figure \ref{fig:Contours2}.



While the einbein action \eqref{einbone} has only one finite pole, intimately related to the presence of a source, there may be many such poles in general.  In fact, it will become clear that presence of other poles is related to the existence of cusp caustics, absent in the present example which has only a smooth caustic.
For the cases discussed in \cite{GLefschetz} in which the einbein action can be determined exactly, it has the form,
\begin{align}
\mathbb S = -i\ln(\Psi)=\sum_j \frac{{\cal R}_j(\vec x)}{\Lambda-\mu_j} + Q(\Lambda,\vec x)  
\, ,
\end{align} 
where the singularities of $Q$ in the complex $\Lambda$ plane include logarithmic branch points and the pole at infinity.   
The contour $\Gamma$ may be chosen to connect the poles of $\mathbb S$,  both at finite $\Lambda$ and $\Lambda=\infty$.  It may at first seem that any pole $\Lambda=\mu_j$ must be associated with a source at points where the residue vanishes, ${\cal R}_j(\vec x) =0$. 
This is true if the contour has just one segment bounded at $\Lambda=\mu_j$. 
If the contour $\Gamma$ contains two segments with boundaries at the same pole but opposite orientations, as shown later for an explicit example in figures \ref{fig:incusp} and \ref{fig:outcusp}, then there is no associated source.  Effectively, two source terms of opposite sign cancel each other.  Note that in this case the two connected segments can be combined and moved away from the pole, assuming they lie on the same Reimann sheet\footnote{Multiple Reimann sheets arise from logarithmic terms in the einbein action having branch points coincident with the poles.}.  However the resulting path is generally no longer  steepest descent.   
The locus of vanishing residue, dubbed a ghost source in \cite{GLefschetz}, has no manifest physical significance in this case. Yet the point at which ghost sources and smooth caustics intersect is physical, corresponding to the singular point of a cusp caustic.
The poles associated with the ghost sources are known as `ghost poles'. In addition to their relevance to cusp caustics, the poles of the einbein action are also intimately related to monodromies \cite{GLefschetz}, in which linearly independent solutions of the Helmholtz equation are mapped into each other upon traversing a closed loop in parameter space.

\section{A simple cusp caustic}\label{simplecusp}

\begin{figure}[!h]
\center{
\includegraphics[width= 200pt]{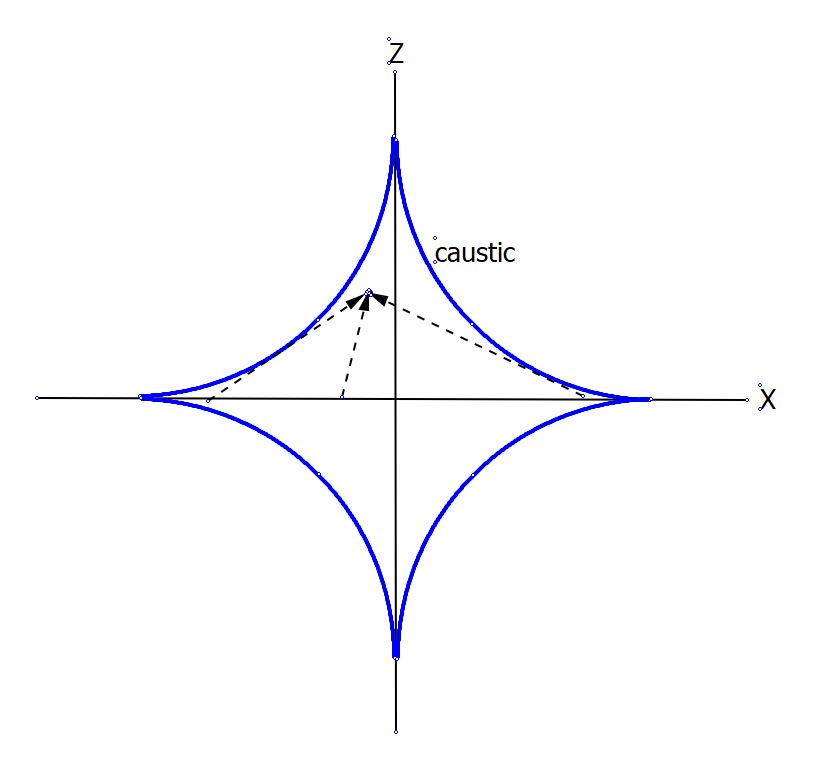}\caption{Three real ray paths for $\vec x$ within the caustic astroid.}
\label{fig:cusprays}
}
\end{figure}

\begin{figure}[!h]
\center{
\includegraphics[width= 200pt]{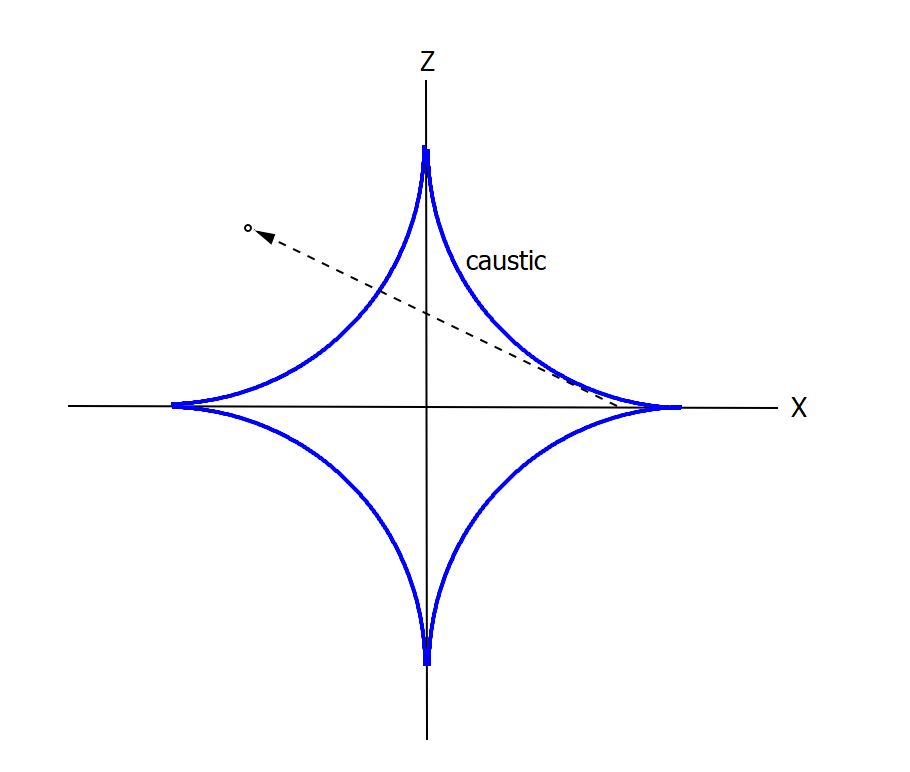}
\caption{Single real ray path for $\vec x$ outside the caustic astroid.}
\label{fig:cusprayout}
}
\end{figure}

A simple example of a cusp caustic for which the einbein action is known exactly is given by the field due to an extended line source with a quadratic phase, 
\begin{align}\label{brf}
J(\vec x) = &\delta(z)exp\left(-ik_0\frac{x^2}{4\mu}\right)\, ,
\end{align} 
where the index of refraction is constant, $n(\vec x)=n_0$.  
The  caustic surface is an astroid given by  
\begin{align}
x^{2/3} + z^{2/3} = \left(2n_0\mu\right)^{2/3}\, ,
\end{align}
with cusp singularities at $(x,z)= (0, \pm 2n_0\mu)$.
The unit vector describing the launch angle of rays at the $z=0$ source is 
\begin{align}
[\hat n_x,\hat n_z] = \left[ -\frac{X}{2\mu},\,\, \sqrt{1-\left( \frac{X}{2\mu}\right)^2} \right] \, ,
\end{align}
such that three rays reach any point inside the cusp, $x^{2/3} + z^{2/3} < (2\mu n_0)^{2/3}$, as shown in figure \ref{fig:cusprays}, while only one real ray reaches points outside, as shown in figure \ref{fig:cusprayout}. 

The solution of the Helmholtz equation with source \eqref{brf} can be written as 
\begin{align}
\phi(\vec x)=\int_0^{+\infty} &d\Lambda \exp\left[ik_0 \bar S(\Lambda,\vec x) + f(\Lambda) \right]  \label{CuspInt} \\
 &\bar S =
 \frac{1}{4(\Lambda- \mu)} x^2 +\frac{1}{4\Lambda}z^2  +\Lambda n_0^2  \label{barsimp}\\
 &f =   \frac{1}{2}\ln\left( \frac{ i\mu}{4\pi k_0\Lambda(\Lambda-  \mu)}\right)
\end{align}
In this and other exactly soluble cases, the einbein action $\mathbb S$ can be separated into a meromorphic and a logarithmic term, $\mathbb S = ik_0\bar S + f$.  We will often refer to the meromorphic term $\bar S$  on its own as the einbein action.  The logarithmic branch points in $f$  coincide with the finite poles of $\bar S$.  The derivation of \eqref{CuspInt} can be found in \cite{GLefschetz}.  

\begin{figure}[!h]
\center{
\includegraphics[width= 200pt]{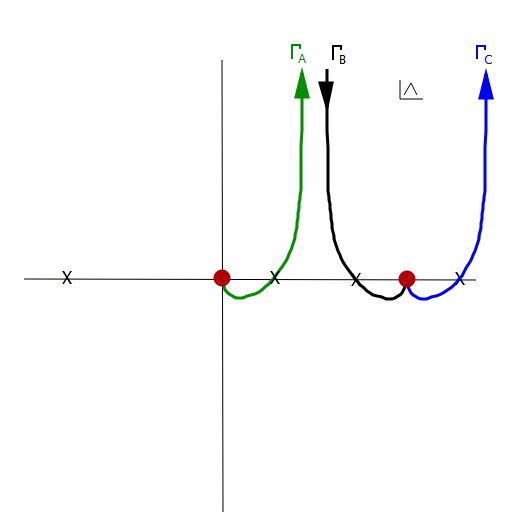}
\caption{Poles, critical points and steepest descent paths for $\vec x$ within the caustic astroid.  Poles are indicated by circles and critical points by X.  The pole on the right is a ghost pole at $\Lambda=\mu$ with residue $\frac{1}{4}x^2$, bounding two steepest descent paths. Each of the paths marked $\Gamma_A,\Gamma_B,\Gamma_C$ correspond to the real rays shown in figure \ref{fig:cusprays}. The path $\Gamma_A$ maps to the ray which does not intersect the caustic. The bounding pole of $\Gamma_A$ at $\Lambda=0$, with residue $z^2/4$ accounts for the source term in the Helmholtz--Green's function equation.}
\label{fig:incusp}
}
\end{figure}

\begin{figure}[!h]
\center{
\includegraphics[width= 200pt]{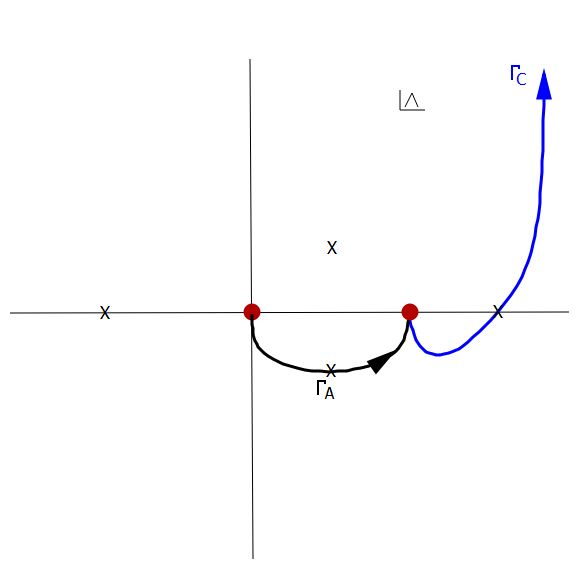}
\caption{Poles, critical points and steepest descent paths contributing to the solution for $\vec x$ outside the caustic astroid.  Poles are indicated by circles and critical points by X.  The pole on the right is a ghost pole, bounding two steepest descent paths. The path $\Gamma_C$ corresponds to the real ray shown in figure \ref{fig:cusprayout}. The path $\Gamma_A$ maps to a complex ray.}
\label{fig:outcusp}
}
\end{figure}

The large $k_0$ limit is controlled by the critical points of $\bar S$, at which $\frac{d\bar S} {d\Lambda} =0$.  At a generic position $\vec x$, the einbein action \eqref{barsimp} has four critical points in the complex $\Lambda$ plane. These are all real within the caustic astroid, $x^{2/3} + z^{2/3} < \left(2n_0\mu\right)^{2/3}$.  The integration contour $\Lambda=[0,+\infty]$ can be deformed into a sum over three steepest descent paths shown in figure \ref{fig:incusp}.  Each of these paths passes through a real critical point corresponding to one of the three real rays which exist inside the caustic astroid.   These steepest descent paths are bounded by the poles $\Lambda=0,\mu,\infty$.  The fourth critical point is a spectator in this case,  through which none of the steepest descent paths comprising $\Gamma$ pass. Outside the astroid, $x^{2/3} + z^{2/3} > \left(2n_0\mu\right)^{2/3}$, two of the critical points become complex.  The contour $\Lambda=[0,+\infty]$ is then equivalent to a sum over the two steepest descent paths shown in figure \ref{fig:outcusp}. One of these paths connects the two finite poles $\Lambda=0$ and $\Lambda=\mu$ passing through a complex critical point. The second path connects the poles $\Lambda=\mu$ and $\Lambda=\infty$, passing through the real critical point corresponding to the single real ray outside the caustic astroid.  The utility of representing solutions in terms of complexified integration contours bounded by essential singularities of the integrand is discussed at length in \cite{GLefschetz}, and is analogous to similar constructions in the context of quantum field theory in \cite{GG1,GG2,Ferrante:2013hg,WIT,Dunne:2015eaa,Behtash:2015loa ,Pehlevan:2007eq,Guralnik:2009pk,Cristoforetti:2012su}.

The residue of the pole of $\bar S$ at $\Lambda=0$ is ${\cal R}=z^2/4$, vanishing along the line $z=0$  corresponding to the source location.  The residue of the pole at $\Lambda=\mu$ is ${\cal R}=x^2/4$, vanishing along the line $x=0$.  Since the representation of the integration contour in terms of steepest descent paths has two components bounded with opposite orientations at this pole, as can be seen in both figures \ref{fig:incusp} and \ref{fig:outcusp}, there is no additional source term in the Helmholtz equation.  The line $x=0$ is instead referred to as a ghost source.  Note that the cusps $x=0,z=\pm 2n_0\mu$ lie along the ghost source $x=0$, with $z$ location along the ghost source determined by the ghost pole $\mu$ and the index of refraction $n_0$.  

It turns out that this cusp localization phenomenon generalizes.  Assuming the validity of the conjectures posed in \cite{GLefschetz},  cusp caustics always lie along ghost sources. Approaching a cusp caustic,  three critical points of the einbein action converge upon a ghost pole $\Lambda \rightarrow \Lambda_p$, such that the pole term $\zeta^2/(\Lambda-\Lambda_p)$ in the einbein action remains finite.   This is physically required since the real component of the einbein action at a critical point is proportional to the physical arrival time \cite{GLefschetz}.   There is also evidence, seen in other examples described in the subsequent sections, that the location of cusp caustics along ghost sources is linearly related to the location of the ghost pole in the $\Lambda$ plane, where it should be noted that $\Lambda$ has dimensions of length.  A precise statement of this relation will be proposed in section \ref{perttheory}.

\section{Cusps and ghost poles for quadratic $n^2$}\label{polepairs}

\begin{figure}[!h]
	\center{
		\includegraphics[width= 420pt]{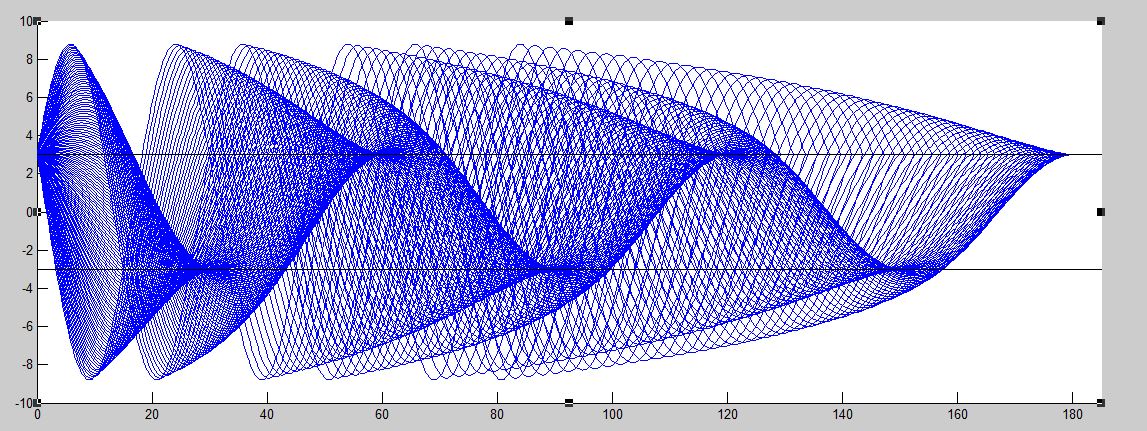}
		\caption{Fan of rays for  the index of refraction $n^2=n_0^2-\alpha Z^2$, with  $\alpha=0.01, n_0=1$ and source at $x_s=0,z_s=3$.  The rays are generated from a set of evenly spaced launch angles with respect to the x axis, between $\pm \pi/3$.  The caustics are manifest as surfaces where the rays become dense.  The cusp singularities occur exactly at the location of ghost sources, $z=\pm z_s$, shown as horizontal lines overlaying the fan of rays.}
		\label{fig:RayFan}
	}
\end{figure}

The previous section demonstrated the presence of cusps for a simple model with a line source having quadratic phase and a constant index of refraction, for which the einbein action has precisely two finite poles.  More generally the einbein action has many poles. In the following discussion, we consider a point source and a quadratic index of refraction
\begin{align}
n(X,Z)^2 &= n_0^2  -\alpha Z^2 \label{chanindex} \\
J&=\delta(\vec x -\vec x_s)\, .
\end{align}
The einbein action is exactly computable \cite{GLefschetz,Palmer,Schlottmann}, given by  
\begin{align}
{\bf \bar{\cal S}}(\Lambda) &= \frac{ (x-x_s)^2  }{4\Lambda} + \Lambda n_0^2 + \sqrt{\alpha} 
\frac{ (z_s^2 + z^2)\cos{ ( 2\sqrt{\alpha}\Lambda) } - 2 z_s z }{2\sin{ (2\sqrt{\alpha}\Lambda )}} \nonumber\\
&= \frac{ (x-x_s)^2  }{4\Lambda} + \Lambda n_0^2 + \frac{\sqrt{\alpha}}{2}(z^2 + z_s^2)\sum_{m=-\infty}^{\infty} \frac{1}{2\sqrt{\alpha}\Lambda - \pi m} - zz_s
\frac{1}{2\Lambda}\prod_{m\ne 0}\frac{\pi m}{\pi m - 2\sqrt{\alpha}\Lambda}\, .\label{chanac}
\end{align}
The residue of the pole at $\Lambda=0$ is 
\begin{align}
{\cal R}_{0} = \frac{1}{4}\left( (x-x_s)^2 + (z-z_s)^2 \right)\, ,
\end{align}
which vanishes at the location of the source.  The other poles are ghost poles, 
\begin{align}
\Lambda= \frac{\pi m}{2\sqrt{\alpha}}
\end{align}
for integer non-zero $m$, having
residues
\begin{align}
{\cal R}_{m} = \frac{1}{4}\left(z- (-1)^m z_s\right)^2\, .
\end{align}
These residues vanish along the the ghost sources $z - (-1)^m z_s =0$.
All cusp caustics lie along the ghost sources, as can be seen in figure \ref{fig:RayFan}. Moving $\vec x$ towards a cusp is accompanied by three critical points converging on a pole of the einbein action.  This behavior differs somewhat from that of the catastrophe generating function for a cusp, $\lambda^4 + \zeta_2 \lambda^2 + \zeta_1 \lambda$, in which case three critical points coalesce at a regular point in the complex $\lambda$ plane as one approaches the cusp at $\zeta_1=\zeta_2=0$. The difference in behavior is reflected in a singularity of the map $\lambda\leftrightarrow\Lambda$ at the cusp.

In this particular example, the position of the cusps along the ghost sources is determined with high accuracy by considering effective einbein actions $\bar S_{\{m\}}$ associated with each ghost pole, in which all other ghost poles in \eqref{chanac} are removed and the index of refraction is replaced with its value at the relevant ghost source: 
\begin{align}\label{polepaircrude}
\bar S_{\{ m\}} = \frac{(x-x_s)^2 + (z-z_s)^2}{4\Lambda} + \frac{ (z-(-1)^m z_s)^2}{4(\Lambda - \frac{\pi m}{2\sqrt{\alpha}})} + n\left((-1)^m z_s\right)^2\Lambda \, ,
\end{align}
analogous to the single ghost pole example of section \ref{simplecusp}.
This yields cusps at
\begin{align}
z&-(-1)^m z_s = 0 \nonumber\\
&\sqrt{(x-x_s)^2 + (z-z_s)^2}= 2 n \left( (-1)^m z_s \right) \frac{\pi m}{2\sqrt{\alpha}}\, ,
\end{align}
in extremely close agreement with the location of the cusps observed by ray tracing.
Although the the statement that cusps are bound to ghost sources is robust,  the linear relation between the ghost poles $\Lambda_m$ and the location of cusps along ghost sources, if  valid in the general case, lacks a concise mathematical statement and explanation, although a proposal is made in section \ref{perttheory}.

\begin{figure}[!h]
	\center{
		\includegraphics[width= 200pt]{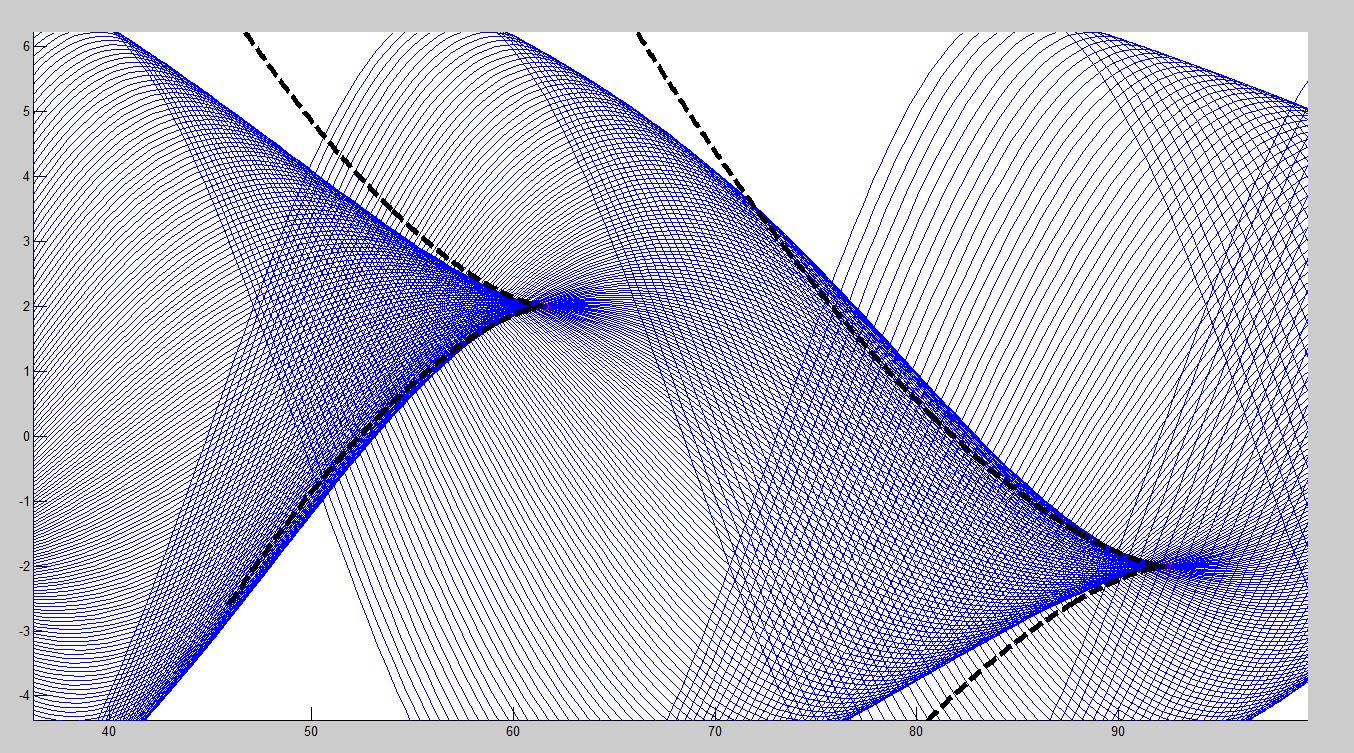}
		\caption{Caustic curves (dashed lines) derived from the effective einbein action \eqref{polepaircrude}, plotted over the ray fan. The position of the cusp singularities is accurately described.  The geometry of the smooth component of the caustic near the  cusp is not as accurately captured until position dependence of the linear term in the effective einbein action is included, as in \eqref{asym}  and figure \ref{fig:Corrected} }
		\label{fig:CuspCrude}
	}
\end{figure}

\begin{figure}[!h]
\center{
\includegraphics[width= 400pt]{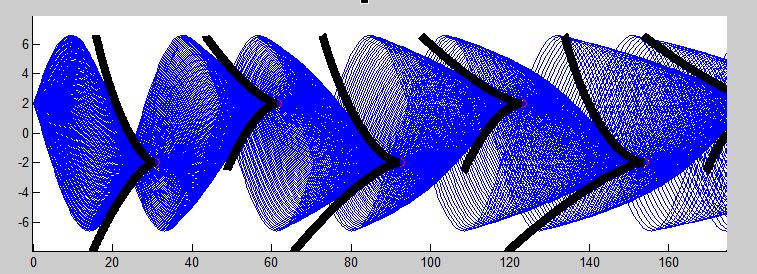}
\caption{Caustic curves (dashed lines) derived from effective pole pair einbein action \eqref{polepaircrude}, plotted over the ray fan. In this case the asymmetry of  the cusps with respect to reflection about the ghost source is captured by the inclusion of position dependence of the term linear in $\Lambda$; $n_{eff} = n( (-1)^m z_s) -B(z-(-1)^m z_s)$}
\label{fig:Corrected}
}
\end{figure}

The approximate caustics derived from \eqref{polepaircrude} are shown overlaying the ray fan in figure \ref{fig:CuspCrude}.  Although the large scale geometry of the smooth components of the caustics is not captured, the smooth caustics in the neighborhood of, and bounded by, the cusp are crudely if not precisely approximated.  The simple approximation \eqref{polepaircrude} yields cusps lacking the asymmetry with respect to reflection about the ghost source axis which is apparent from the ray trace shown in figure \ref{fig:CuspCrude}.  However there is no reason to expect that the effective index of refraction in \eqref{polepaircrude} should be position independent.  For the effective two pole action  
\begin{align}
\bar S_{\{m\}} \approx k_0\left[\frac{r_1(\vec x)^2}{4\Lambda} + \frac{r_2(\vec x)^2}{4(\Lambda-\mu)} + n_{eff}^2(r_1,r_2)\Lambda\right]
\end{align} 
the caustic has the form of the corner of a deformed  astroid,
\begin{align}
r_1^{2/3} + r_2^{2/3} = (2n_{eff}(r_1,r_2)\mu)^{2/3}\, .
\end{align}
Including an asymmetric dependence of $n_{eff}$ with respect to reflection about a ghost source $z=z_g$, such as 
\begin{align}\label{asym}
n_{eff} = n( (-1)^m z_s) -B(z-(-1)^m z_s)\, .
\end{align}
with a suitably tuned parameter $B$
yields a much better match to the geometry of smooth caustics in the immediate neighborhood of the cusp, as shown in figure \ref{fig:Corrected}.
Note that in the present example a $z$ dependent perturbation of $n_{eff}$ deforms the caustic curve without moving the cusp singularity, while an   
$x$ dependent perturbation slides the cusp singularity along the ghost source.

The effective einbein action involving a single ghost pole is akin to a uniform asymptotic approximation associated with a particular cusp caustic.  
Approximations to the einbein action necessarily preserve the behavior in the neighborhood of selected critical points, but need not preserve the steepest descent paths and their boundaries at either infinity or other poles in the complex $\Lambda$ plane.  In particular there is no obvious reason why the source pole in the effective model need be identical to that of the full solution, although it seems to be equivalent in the present example. 
In the subsequent section, we discuss the map between the single ghost pole effective einbein action and the catastrophe generating function which enters the integral representation of the Pearcey function, arising in a uniform asymptotic description of the cusp caustic.

\section{Mapping spatial coordinates to the arguments of the Pearcey function}\label{Pearceymap}
 
After a suitable transformation $\lambda(\Lambda,\vec x)$ for $\vec x$ in the neighborhood of a particular caustic, the einbein action formulation of the solution of the Helmholtz equation is equivalent to a uniform asymptotic approximation\cite{Pearcey,Chester,Kravtsov1,Kravtsov2,Ludwig,Berry,Berry2,Berry3}.  In this approximation,  the field is written as an integral 
\begin{align}\label{intrep}
\Phi(\vec x) \approx \int_{-\infty}^{\infty} d\lambda f(\vec x,\lambda) e^{ik_0S(\lambda,\vec x)}\, . 
\end{align} 
where the function $S$ is a polynomial in $\lambda$, also known as the catastrophe generating function. There are different polynomials associated with each type of caustic. For a smooth or fold caustic, 
\begin{align} \label{smoothpoly}
k_0\bar S \approx \lambda^3 + \zeta\lambda(\vec x)
\end{align}
such that \eqref{intrep} can be evaluated in terms of the Airy function $Ai(\zeta)$ and its derivatives, whereas for a cusp caustic,
\begin{align}\label{cusppoly}
k_0\bar S \approx \lambda^4 + \zeta_2(\vec x)\lambda^2 + \zeta_1(\vec x)\lambda
\end{align}
in which case \eqref{intrep} is expressed in terms of the Pearcey function $P(\zeta_2,\zeta_1)$.
The challenge in either case is to obtain the relation between the spatial coordinates $\vec x$ and the arguments $\zeta_I$. 

The einbein action $\bar S(\Lambda)$ and the generating functions $G(\lambda)$ for $A_n$ catastrophes are similar in that both are functions of a single variable, which when exponentiated and integrated yield the solution of the Helmholtz equation.  However the catastrophe generating functions yields a uniform asymptotic approximation associated with the neighborhood of particular caustic, whereas the einbein action contains complete information about the field including the entire caustic web.
If one wishes to describe the neighborhood of a single cusp only,  then the single ghost pole model 
suffices, having the general form
\begin{align}\label{effein}
\bar S \approx k_0\left[\frac{r_1(\vec x)^2}{4\Lambda} + \frac{r_2(\vec x)^2}{4(\Lambda-\mu)} + n_{eff}^2(r_1,r_2)\Lambda\right],
\end{align}
where the source at $r_1(\vec x)=0$ is not necessarily the physical source as noted in the previous section.

As discussed in \cite{GLefschetz}, the Schr\"oedinger equation \eqref{Schrod} implies that the residues ${\cal R}(\vec x)$ satisfy 
\begin{align}
-(\vec\nabla {\cal R})^2 +  {\cal R} = 0\\
\vec\nabla^2{\cal R} - \frac{D}{2} =0\, ,
\end{align}
where the integer $D$ is the codimension of the surface on which ${\cal R}=0$.  If follows that $r_1$ and $r_2$ in \eqref{effein} are distance to surfaces of codimension $D_{1,2}$. The effective einbein action \eqref{effein} is analogous to the cusp generating function \eqref{cusppoly},  but is formulated directly in terms of spatial coordinates and has a different analytic structure.
 
The parameters of the catastrophe generating function depend on  spatial coordinates in a way which is generally obtained by comparing with geometric optics results at points away from the caustic \cite{Ludwig}.  Given the einbein description of a cusp caustic, this dependence can also be constructed from the  map $\Lambda \leftrightarrow \lambda$ between the single ghost pole  effective action \eqref{effein} and the cusp catastrophe generating function, such that
\begin{align}\label{comptwo}
\bar S \approx \lambda^4 + \zeta_2 \lambda^2 + \zeta_1 \lambda + \Phi(\zeta_1,\zeta_2) \approx k_0\left[\frac{r_1(\vec x)^2}{4\Lambda} + \frac{r_2(\vec x)^2}{4(\Lambda-\mu)} + n_{eff}^2(r_1,r_2)\Lambda\right]\, .
\end{align} 
The map is complicated by the fact that the einbein action has pole singularities whereas the generating function is a polynomial.  In the vicinity of the fold caustics bound by the cusp, the map takes the form 
\begin{align}
\lambda = A(\vec x)(\Lambda-B(\vec x)) + 
\cdots 
\end{align}
and is obtained by comparing the cusp generating function with a truncated Taylor-Maclauren expansion of the einbein action.
Such a map is necessarily singular at the cusp; approaching a cusp in the einbein description involves three critical points converging on a pole in the $\Lambda$ plane, whereas in the catastrophe generating function three critical points coalesce at a regular point in $\lambda$.   In the subsequent discussion, the map between the einbein action and and the catastrophe generating function for a cusp is constructed, yielding the relation between  the spatial coordinates and the arguments of the Pearcey function.

\begin{figure}[!h]
\center{
\includegraphics[width= 300pt]{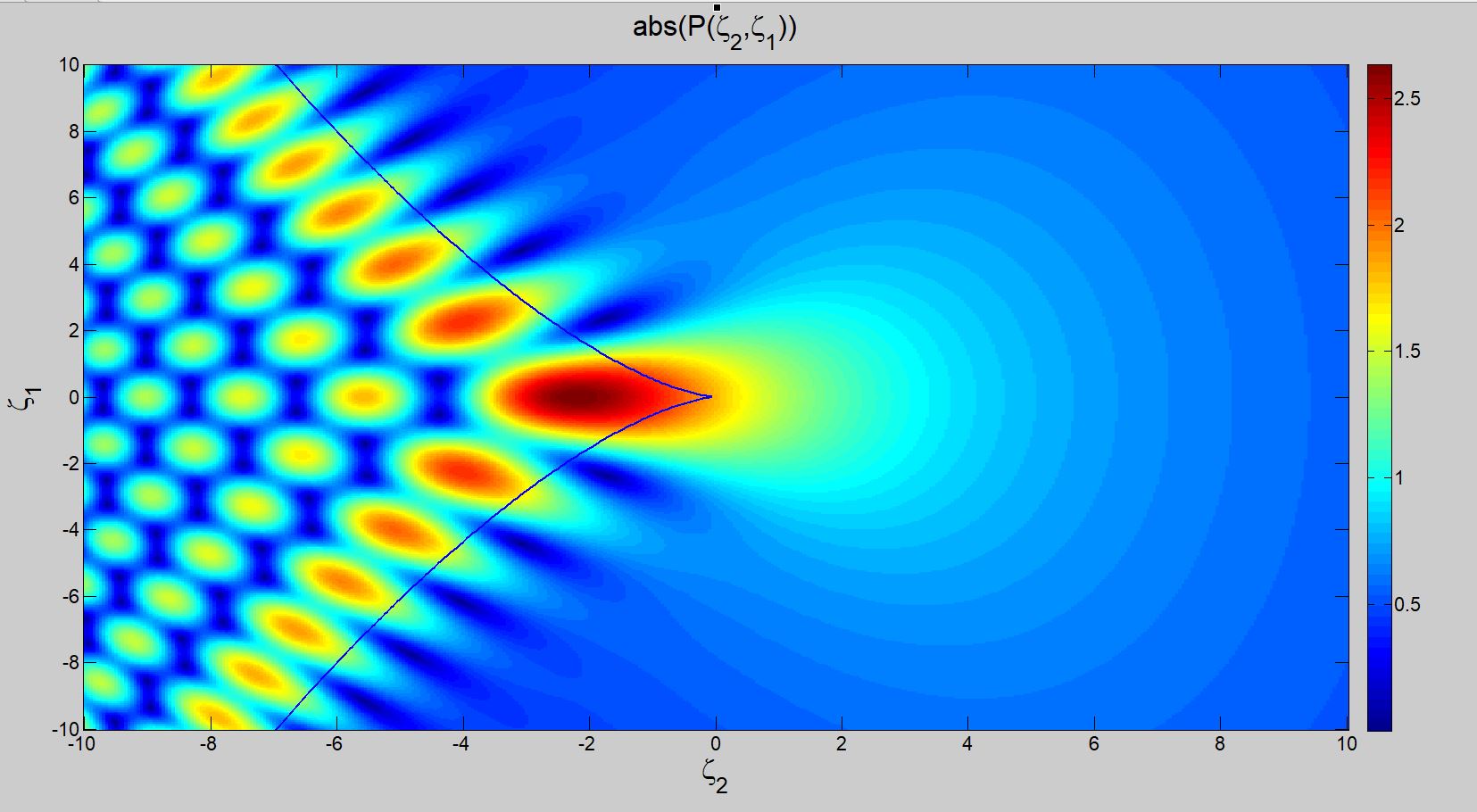}
\caption{Absolute value of the Pearcey function in the neighborhood of a cusp. There is a non-trivial map between the arguments $\zeta_{1,2}$ and spatial coordinates $\vec x$,  which is readily determined using the einbein formulation.}
\label{fig:PearceyFn}
}
\end{figure}



The caustic curve is obtained by requiring vanishing first and second derivatives of the einbein action \eqref{effein} with respect to $\Lambda$, giving
\begin{align}
r_1^{2/3} + r_2^{2/3} = (2n_{eff}\mu)^{2/3}\, .
\end{align}
Expanding about the cusp $r_1=2n_{eff}\mu$ by defining $\tilde r_1 \equiv r_1 - 2n_{eff}\mu$, this becomes
\begin{align}
\left(\frac{\tilde r_1}{2n_{eff}\mu}\right)^3 + \frac{27}{8}r_2^2=0
\end{align}
in the limit $\tilde r_1<<2n_{eff}\mu << 1,\,\,r_2 << 2n_{eff}\mu$.  This is to be compared with the caustic curve derived from the vanishing of first and second derivatives of the cusp generating function \eqref{cusppoly} with respect to $\lambda$,
\begin{align}
 \zeta_2^3 + \frac{27}{8}\zeta_1^2=0
\end{align}
Therefore one expects the map 
\begin{align}
\zeta_2 &= \gamma\frac{\tilde r_1}{(2n_{eff}\mu)^{1/3}}\\
\zeta_1 &= \gamma^{3/2}r_2
\end{align}
for some yet to be specified $\gamma$.  To determine $\gamma$ one must compare not just the form of the caustic curve but the two representations of the action in \eqref{comptwo}

Since the map $\Lambda\leftrightarrow\lambda$ is singular at the cusp,  it is easier to compute $\zeta_{I}(\vec x)$ by comparing the generating function and the einbein action in the neighborhood of a smooth component of the caustic.  In other words one has the threefold equivalence,
\begin{align}\label{compthree}
\bar S\approx \lambda^4 + \zeta_2 \lambda^2 + \zeta_1 \lambda + \Phi(\zeta_1,\zeta_2) &\approx k_0\left[\frac{r_1(\vec x)^2}{4\Lambda} + \frac{r_2(\vec x)^2}{4(\Lambda-\mu)} + n_{eff}^2(r_1,r_2)\right] \nonumber\\
\approx \lambda'^3 + \zeta'\lambda' + \Phi'(\zeta')
\end{align}
whereby one maps both the einbein action and cusp generating function to the generating function of a smooth caustic \eqref{smoothpoly}, via
\begin{align}\label{trif}
\lambda'=a(\lambda-b)=c(\Lambda-d)\, .
\end{align}  
The coefficients $a,b,c,d$ are chosen so that $\bar S$ contains no $\lambda'^2$ term and the coefficient of $\lambda'^3$ is one.
The coefficient $\zeta'$ of the $\lambda'$ term must be the same whether one starts with the einbein action or the cusp generating function.
Starting with the einbein action, one finds
\begin{align}\label{zetae}
\zeta' =  &k_0^{3/4}\left(\frac{3}{8}\right)^{1/6}\mu^{-2/3}\,\Omega^{4/3}\,(-\Gamma)^{1/6}\left[\chi - \left(\frac{8}{27}\right)^{1/2}(-\Gamma)^{3/2}\right], 
\end{align}
with 
\begin{align}
\Omega\equiv 2 n_{eff}\mu, \,\,\,\,
\Gamma\equiv \frac{\tilde r_1}{\Omega},\,\,\,\,\,
\chi\equiv \frac{r_2}{\Omega}\, ,
\end{align}
where proximity to the cusp singularity means $\Gamma<<1$ and $\chi<<1$.
Starting with the cusp generating function gives 
\begin{align}\label{zetag}
\zeta'=4^{-1/3}\,6^{1/6}\,(-\zeta_2)^{-1/6}\left[\zeta_1 - \left(\frac{8}{27}\right)^{1/2}(-\zeta_2)^{3/2}\right]
\end{align}
Comparing \eqref{zetae} and \eqref{zetag} gives the map between arguments of the generating function and spatial coordinates,
\begin{align}\label{themap}
\zeta_2 = k_0^{1/2}\mu^{-1/2}(r_1 - 2n_{eff}\mu)\nonumber\\
\zeta_1 = k_0^{3/4}\mu^{-3/4}(2n_{eff}\mu)^{1/2}r_2\, .
\end{align}
Similiar arguments yield the phase $\Phi$ in the relation \eqref{comptwo}
\begin{align}
\Phi = ik_0 \frac{\Omega^2}{\mu}\left(\frac{3}{4}\chi^{2/3} + 2 \Gamma\right)
\end{align}

The map between $\lambda$ and $\Lambda$, following from \eqref{trif}, 
has the form $\lambda = \frac{c}{a}(\Lambda-d) + b$ where 
\begin{align}
\frac{c}{a} \sim \zeta_2^{-1/6}\chi^{-2/9}
\end{align}
which is singular along the curve $\chi=0$, the locus of the ghost source, and also along the curve $\zeta_2=0$. These two curves intersect at the cusp singularity. The singularity of the map is to be expected, since the catastrophe generating function is polynomial whereas the einbein action has poles, with the latter playing a special role in the localization of cusp caustics.


\section{The origin of poles and ghost sources} \label{origin}
 

The singularities of the einbein action have origins in a Dirichlet problem related to the Feynman path integral formulation, 
\begin{align}
e^{i\mathbb S(\Lambda,\vec x,\vec x_s)} &=  \int D\vec X(\tau)\exp\left(ik_0 \int_0^1 d\tau {\cal L}\left(\vec X,\frac{d\vec X}{d\tau}\right)\right)  \label{PI} 
\end{align}
with Lagrangian
\begin{align}
{\cal L}\equiv\frac{1}{4\Lambda}\left( \frac{d\vec X}{d\tau} \right)^2   + \Lambda n^2(\vec X)\, , \label{Lagr}  
\end{align}
where one integrates over all paths on the interval $\tau=[0,1]$ having endpoints  $X(0)=\vec x_s$ and $X(1)=\vec x$ corresponding to the arguments of the Green's function.  
A pole of the einbein action $\mathbb S$ at $\Lambda=0$ is already apparent.  Other poles arise upon carrying out the path integration over $\vec X(\tau)$.  These  poles correspond to $\Lambda$ at which the Euler-Lagrange equations,
\begin{align}
\frac{d}{d\tau}\left(\frac{\partial {\cal L}}{\partial \dot{\vec X}}\right) - \frac{\partial {\cal L}}{\partial \vec X} = 0,
\end{align}
have no solution for generic Dirichlet boundary conditions. In this case, a solution exists only for special values of $\vec x$ comprising the ghost sources, or points at which the residue of the ghost poles vanish. 

As an example, for the index of refraction given in \eqref{chanindex}, the Euler-Lagrange equations derived from $\cal L$ are
\begin{align}
\frac{1}{2\Lambda}\frac{d^2 X}{d\tau^2}& = 0\nonumber\\
\frac{1}{2\Lambda}\frac{d^2 Z}{d\tau^2}& + 2\Lambda \alpha Z = 0\, , 
\end{align} 
having solution
\begin{align} \label{Lamsoln}
X&=A+B\tau\nonumber\\
Z&=C \cos(2\sqrt{\alpha}\Lambda\tau + \Theta)\, .
\end{align}
Subject to Dirichlet boundary conditions at $\tau=0,1$, these equations have no solution for $2\sqrt{\alpha}\Lambda=\pi m$ unless $Z(1) = (-1)^m Z(0)$, corresponding to the ghost poles and ghost source of the einbein action \eqref{chanac}.  It must be emphasized that the trajectories satisfying the Euler Lagrange equations derived from ${\cal L}$ in \eqref{Lagr} are not the same as rays. Ray theory arises from a different order of integration in the path integral representation of the Green's function, in which integration over an einbein  field $\Lambda(\tau)$ is carried out before that over $\vec X(\tau)$, as explained in \cite{GLefschetz}. The  stationary phase paths in $\vec X$ are then the standard rays satisfying the Eikonal constraint, for which $\Lambda(\tau)$ is a Lagrange multiplier.

For fixed Dirichlet boundary conditions with endpoint $\vec X(1)$ away from any ghost source,  the solutions of the Euler Lagrange equations diverge as $\Lambda$ approaches a ghost pole.  For example, the solutions \eqref{Lamsoln} have $C\rightarrow\infty$ as $2\sqrt{\alpha}\Lambda\rightarrow \pi m$ for fixed Dirichlet boundary conditions.  The divergence of  solutions of the Dirichlet problem is of particular importance when the effect of  perturbations of the index of refraction on ghost poles is considered, as discussed in the next section.    

\begin{figure}[!h]
\center{
\includegraphics[width= 300pt]{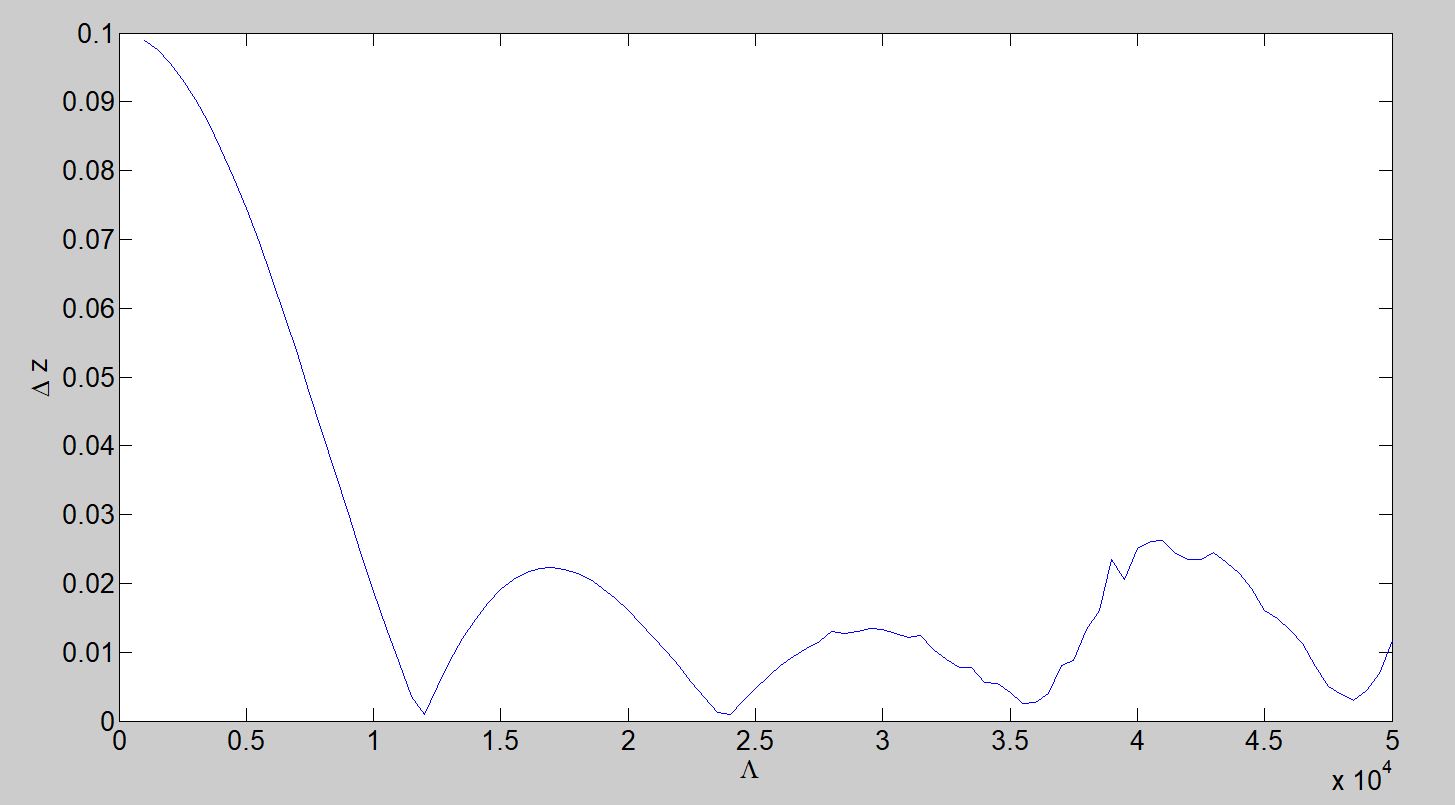}
\caption{The difference between the maximum and minimum endpoints $z(1)$ as a function of $\Lambda$, for some spread of initial conditions $\dot z(0)$ in the case of the Munk profile \eqref{MunkProfile}  with $\epsilon = 0.00737,\, z_c=1300$ and a source at $z=1340$.  The endpoints are found by solving the Euler Lagrange equations derived from the Lagrangian \eqref{Lagr}, which are not to be confused with the equations of ray theory.
The values of $\Lambda$ at which the  spread of endpoints collapses correspond to singularities in the einbein action.}
\label{fig:MunkLambdaFan}
}
\end{figure}

\begin{figure}[!h]
\center{
\includegraphics[width= 300pt]{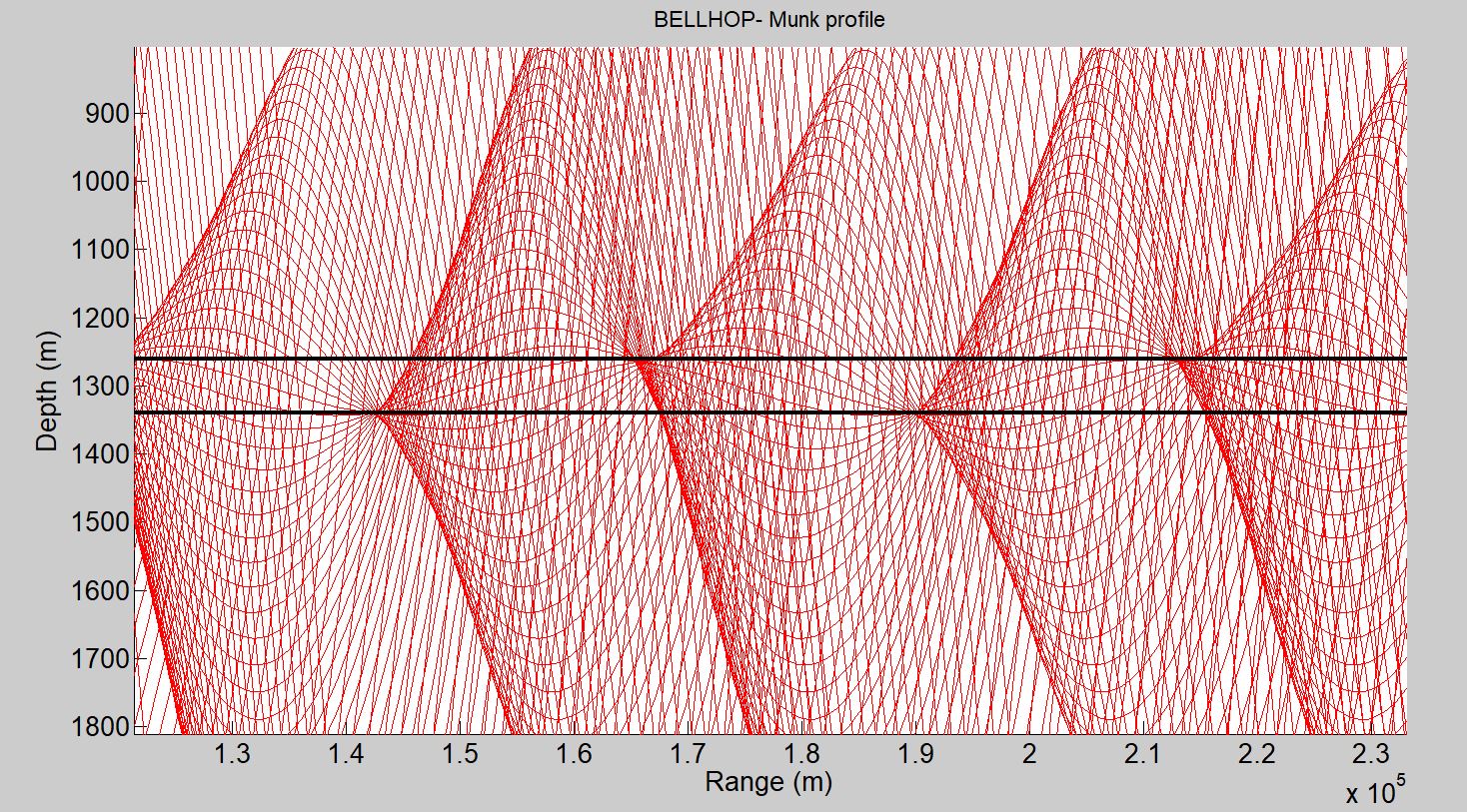}
\caption{Ray fan for the Munk profile \eqref{MunkProfile} with $\epsilon = 0.00737,\, z_c=1300$ and a source at $z=1340$, computed numerically using the gaussian beam code `Bellhop' \cite{Bhop1,Bhop2}. The superimposed horizontal lines are the ghost sources, computed numerically by means described in section \ref{origin}. Cusps lie precisely along the ghost sources.  }
\label{fig:MunkRayFan}
}
\end{figure}

When the index of refraction is such that an exact solution for ${\mathbb S}$ is not known, one can hunt for singularities by solving the Euler Lagrange equations numerically and looking for $\Lambda$ at which the map between the initial velocities $\dot {\vec X}(0)$ for fixed $\vec X(0)$ and the endpoints $\vec X(1)$ degenerates.  Figure \ref{fig:MunkLambdaFan} shows this analysis for the two dimensional case with a Munk profile \cite{Munk},
\begin{align}\label{MunkProfile}
n(z)=\frac{1}{ 1+\epsilon\left( 2\frac{z-z_c}{z_c} - 1 + e^{-2\frac{z-z_c}{z_c}}\right)}\, .
\end{align}
As $\Lambda$ is varied for some range of initial velocities, the spread of endpoints $z(1)$  collapses to a single point (a ghost source) at special values of $\Lambda$ (a ghost pole).  Figure \ref{fig:MunkRayFan} shows the ghost sources computed in this manner, superimposed on a fan of rays for the Munk profile.  The cusps are clearly seen to lie along the ghost sources. 

\begin{figure}[!h]
\center{
\includegraphics[width= 400pt]{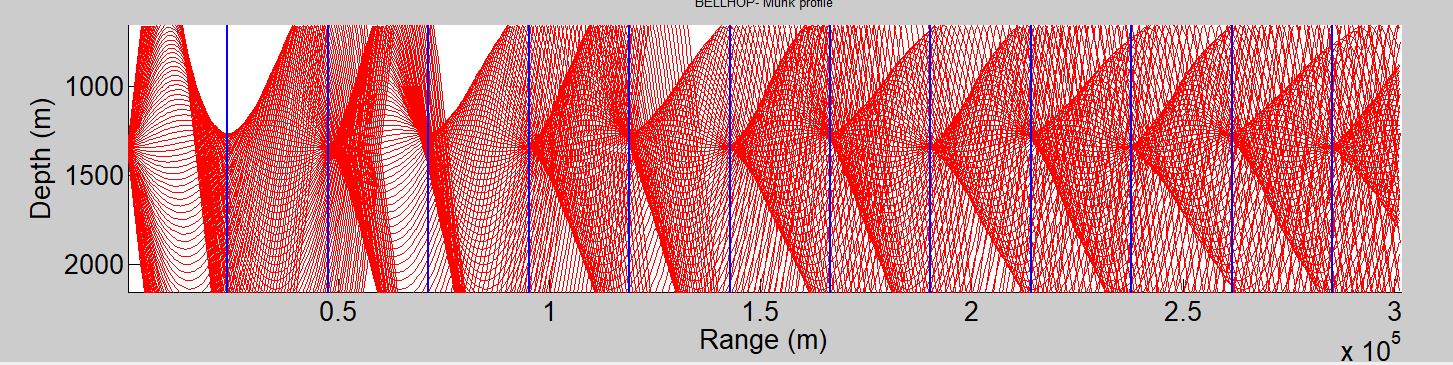}
\caption{Ray fan for the Munk profile \eqref{MunkProfile} with $\epsilon = 0.00737,\, z_c=1300$ and a source at $z=1340$, computed using Bellhop \cite{Bhop1,Bhop2}. The cusp ranges  $x=2n_{eff}\Lambda_m$, estimated from a numerical computation of the ghost poles, is superimposed.}
\label{fig:MunkFanLambdaRanges}
}
\end{figure}

For the Munk profile, the location of cusps along the ghost sources is captured to surprising accuracy by $x=2 n_{eff}\Lambda_m$, where $\Lambda_m$ are the poles determined numerically by the collapse of the Dirichlet problem and $n_{eff}$ is the index of refraction at the ghost source depth, as shown in figure \ref{fig:MunkFanLambdaRanges}. This is very similar to the result for the exactly soluble case \eqref{chanindex}, with cusp locations captured by the single ghost pole effective actions \eqref{polepaircrude}. 

A repetition of the above analysis for an index of refraction depending on more than a single variable would be very interesting, but has yet to be attempted.  The Euler-Lagrange equations for ${\cal L}(\vec x(\tau),\Lambda)$ in the $x$ independent case described here imply that the ghost sources are lines of constant $z$. Ghost sources will presumably be curved in the more general case.

The localization of cusps observed here is a special case of a more general relation proposed in section \ref{perttheory}.
It will be shown in the next section that the einbein action may also have essential singularities in the cases in which $n^2$ is not quadratic.  
We have assumed that the collapse of the Dirichlet problem for the Munk profile corresponds to poles.  It will be argued that essential singularities of ${\mathbb S}(\Lambda)$ correspond to a partial rather than complete collapse of the Dirichlet problem and are related to the higher order  $A_4$ (swallowtail) and $A_5$ (butterfly) caustics. Numerical evidence will be given suggesting that the location of these caustics is also related to degenerate points of the Dirichlet problem in the space spanned by $\Lambda$ and the endpoint boundary conditions.

\section{Perturbations, einbein singularities and higher order caustics}\label{perttheory}

For cases in which the exact einbein action is known, or quadratic $n(\vec x)^2$, the singularities are poles and the highest order caustics are cusps. In the following, we consider the effect of perturbations on exactly soluble cases.  One result of this analysis will be that the einbein action may also include essential singularities in $\Lambda$. Comparing with numerical ray tracing, we will extend the conjecture relating poles and cusps to include a relation between essential singularities and higher order caustics, giving an explicit example for an $A_5$ or butterfly caustic.
Essential singularities differ from poles in that the einbein action does not diverge as $\Lambda$ approaches the singularity $\Lambda_p$ within an angular domain including the real axis.  Consequently, the Dirichlet problem does not collapse completely; finite solutions of the Euler Lagrange equations associated with the Lagrangian \eqref{Lagr} can be found for generic boundary conditions at $\Lambda_p$.
In fact the structure of the degeneration of the Dirichlet problem in the $\Lambda$--endpoint space very closely resembles the spatial geometry of the higher order caustic. Another result of the perturbative analysis will be that the singularities, and the associated aspects of caustic geometry, are very robust against a large class of local variations of the index of refraction.

Consider a parameterized deformation of the the sound speed profile,
\begin{align}
n^2(\vec X) = n_0^2(\vec X) + \rho\Omega(\vec X)
\end{align}
so that the new Lagrangian \eqref{Lagr} is
\begin{align}
{\cal L}_\rho = {\cal L}_0 + \rho\Lambda \Omega(\vec X(\tau))\, 
\end{align}
and the path integral which yields the einbein action is
\begin{align}\label{deltaS}
e^{i{\mathbb S}_\rho}= 
\int D\vec X&
e^{	ik_0 \int d\tau 
	 {\cal L}_\rho 
	}\, .
\end{align}
Under variations of $\rho$, the change in the einbein action is proportional to the average value of the perturbation $\int_0^1 d\tau \Omega(\vec X(\tau))$, summed over all paths between endpoints with the weight $\exp(ik_0\int d\tau {\cal L}_\rho)$;
\begin{align}\label{vev}
\frac{\partial}{\partial \rho} &{\mathbb S}_\rho = 
k_0\Lambda\langle\Omega\rangle \nonumber\\
\langle\Omega\rangle &\equiv
k_0\Lambda\frac{
\int D\vec X \left(\int d\tau \Omega(\vec X)\right)
	e^{
	ik_0 \int d\tau 
		 {\cal L}_\rho 
	}
	}
	{
\int D\vec X 
e^{
	ik_0 \int d\tau 
		 {\cal L}_\rho
	}
}\, .
\end{align}
In the large $k_0$ limit, $\langle\Omega\rangle$ is determined by the solution of the Euler Lagrange equations derived from ${\cal L}_\rho$,
\begin{align}\label{expval}
\langle\Omega\rangle = \int d\tau \Omega(\vec X_{cl}(\tau) )
\end{align}

In the previous section, it was observed that solutions of the Euler Lagrange equations derived from  \eqref{Lagr}, with fixed endpoints away from ghost sources,  diverge  as $\Lambda$ approaches poles of ${\mathbb S}(\Lambda)$ at which the Dirichlet problem has no solution. 
Solutions $\vec X_{cl}(\tau)$ which diverge in this limit will evade most perturbations confined to a region of finite $\vec X$, so that \eqref{expval} 
vanishes for $\Lambda$ within some neighborhood of a pole $\Lambda_p$.
When the divergent solutions pass through the perturbed region, they do so in a vanishingly small parameter time $\Delta \tau$ as $\Lambda\rightarrow\Lambda_p$.
Thus \eqref{expval} may only diverge as $\Lambda-\Lambda_p = \delta\rightarrow 0$ for non-local perturbations $\Omega(\vec X)$.  Hence poles and their residues can not change continuously under local variations of the index of refraction. We will describe their motion via \eqref{vev} for some non-local perturbations below,  and then consider local and semi-local variations under which poles become essential singularities.  Although essential singularities are also robust against large classes of local perturbations, they are not immovable in the way poles are.


The structure of ghost poles does generically change under non-local perturbations of the index of refraction. While $\langle\Omega\rangle$, defined by \eqref{vev}, can only be singular for $\Lambda=\Lambda_p$ at which the exponent in the path integral has no saddle point, the singularity can indicate motion of the pole or the ghost source under variation of the perturbation parameter $\rho$. 

One can check that \eqref{expval} correctly reproduces the motion of a ghost ghost pole or ghost source due to variations of the parameters in the exactly soluble example of section \ref{polepairs}.  In this case variation of the parameter $\alpha$, which is a non-local perturbation corresponding to $\Omega=Z^2$, moves the ghost poles which lie at $\Lambda_p = \frac{\pi m}{2\sqrt{\alpha}}$.
In the vicinity of a ghost pole, $\Lambda-\Lambda_p= \delta$, one obtains
\begin{align}\label{dS}
\frac{\partial}{\partial \alpha} {\mathbb S} &= 
k_0\Lambda\int_{0}^{1} d\tau
\Omega(\vec X_{cl}(\Lambda,\tau)) =  k_0\Lambda\int_{0}^{1} d\tau Z_{cl}(\Lambda,\tau)^2 \nonumber \\ &=  k_0\Lambda C^2 \int_{0}^{1} d\tau cos^2(2\sqrt{\alpha}(\Lambda_p+\delta) \tau + \Theta) \nonumber \\ &\approx \frac{1}{2}k_0\Lambda_p C^2 \, ,
\end{align}
where we have used the solution of the Euler Lagrange equations \eqref{Lamsoln}.
The Dirichlet boundary conditions $Z(0)=z,\, Z(1)=z'$ give the parameters $C$ and $\Theta$, with 
\begin{align}\label{C}
C \approx \frac{z-(-1)^m z'}{2\sqrt{\alpha}\delta}
\end{align}
in the limit of small $\delta$. The divergence of $C$ as $\delta\rightarrow 0$ is a reflection of the absence of a solution of the Dirichlet problem for $\Lambda= \Lambda_p$ when $z'$ does not lie on ghost source. Inserting \eqref{C} into \eqref{dS} gives 
\begin{align}
\frac{\partial}{\partial \alpha} {\mathbb S} 
\approx -k_0\frac{1}{2}\Lambda_p\frac{ (z-z')^2 }{4\alpha (\Lambda-\Lambda_p)^2}
=
k_0 \frac{d\Lambda_p}{d \alpha}\frac{\partial}{\partial\Lambda_p}\left(\frac{(z-z')^2}{4(\Lambda-\Lambda_p)}\right)
\end{align}
consistent with translation of the ghost pole.   Another non-local perturbation is given by $\Omega =Z$ and  $n^2 = n_0^2 - \alpha Z^2 + \rho \Omega$.  In this case one can repeat arguments of the type just described, using the behavior of solutions of the Dirichlet problem as $\Lambda-\Lambda_p \rightarrow 0$ to show that $
\partial_\rho {\mathbb S} \approx k_0  \zeta/(\Lambda - \Lambda_p)\sim \partial_\zeta ( \zeta^2/(\Lambda-\Lambda_p))$, consistent with translation of the ghost source. 

Although poles can not move continuously under local perturbations, one can interpolate between poles with semi-local variations. Consider the index of refraction  
\begin{align}
n^2(\vec x) = n_0^2 -(\alpha-\beta) Z^2 -  \beta Z^2 e^{-\rho Z^2}\, .
\end{align}
The perturbation associated with variation of $\rho$ is referred to as semi-local because it is local in $z$ but independent of $x$.
For $\rho = 0$ there are poles at $\Lambda = \frac{\pi m}{2\sqrt{\alpha}}$, whereas for $\rho\rightarrow\infty$ the poles lie at $\Lambda = \frac{\pi m}{2\sqrt{\alpha-\beta}}$.  Yet continuous variation from $\rho=0$ to $\rho=\infty$ can not continuously move the pole.  In this case 
\begin{align}
\Omega = \frac{\partial}{\partial\rho}\left(n^2(\vec X)\right) = \beta Z^4 e^{-\rho Z^2}
\end{align}
so that 
\begin{align}\label{vevloc}
\frac{\partial}{\partial \rho} {\mathbb S} = 
k_0\Lambda \int_0^1 d\tau \beta Z_{cl}(\tau)^4 e^{-\rho Z_{cl}(\tau)^2}
\end{align}
If the pole moved continuously with $\rho$, \eqref{vevloc} would necessarily reflect this motion by being divergent in the neighborhood of the pole.  However the exponential fall off of the perturbation at large $Z$, where classical solutions in the neighborhood of the pole spend all but an infinitesimal amount of time $\tau$, implies that \eqref{vevloc} vanishes as the pole is approached. The resolution of this conundrum is that the pole becomes an essential singularity at intermediate $\rho$.  As $\Lambda$ approaches the singularity, $\langle\Omega\rangle$ initially grows as if there were a pole, but then vanishes rapidly for an approach within some angular domain which includes the real axis.  For the sake of illustration, an example of a function having an essential singularity with this behavior is 
\begin{align}
\frac{1}{\Lambda-\Lambda_m}e^{-\gamma\frac{1}{(\Lambda-\Lambda_m)^2}}\, .
\end{align}
An explicit computation of an essential singularity in $\Lambda$ arising due to a perturbation is described below.

The appearance of an essential singularity in \eqref{vevloc} at finite non-zero $\rho$  somewhat difficult to show explicitly since the Euler Lagrange equations for \eqref{Lagr} lack an analytic solution for non-zero finite $\rho$.  However one can demonstrate the appearance of essential singularities at intermediate values via a different deformation.  Consider instead 
\begin{align}\label{sigdef}
n^2(\vec X) = n_0^2 - \alpha Z^2 - \sigma Z^2 e^{-\rho Z^2}\, ,
\end{align}
with fixed non-zero finite $\rho$, and $\sigma$ taken to be the varying parameter.
For very small but finite $\rho$, ray tracing is scarcely different from the case $\rho=0$,  suggesting that cusps will move in manner consistent with motion of the ghost poles in the case $\rho=0$.  In fact, we shall show that poles are replaced with a essential singularities as $\sigma$ migrates away from zero.
Under variations of $\sigma$, 
\begin{align}
\frac{\partial {\mathbb S}}{\partial \sigma} 
= k_0\Lambda \int_0^1 d\tau  Z_{cl}(\tau)^2 e^{-\mu Z_{cl}(\tau)^2}\, .
\end{align}
For $\Lambda$ near a pole $\Lambda_m$, $Z_{cl}$ is given by \eqref{Lamsoln} with the near pole behavior \eqref{C} such that
\begin{align} \label{vansh}
\left.\frac{\partial {\mathbb S}}{\partial \sigma}\right|_{\sigma=0} &\approx k_0 \Lambda_m \int_0^1 d\tau C^2 cos^2(\pi m \tau )e^{-\rho C^2 cos^2(\pi m \tau)} \nonumber \\ 
&= \frac{1}{2}k_0\Lambda_m C^2e^{-\frac{1}{2}\rho C^2}\left[ I_0\left(\frac{1}{2}\rho C^2\right) - I_0'\left(\frac{1}{2}\rho C^2\right)\right], \nonumber\\
C = &\frac{z-(-1)^m z'}{2\sqrt{\alpha}(\Lambda-\Lambda_m)} \nonumber\\
\Lambda_m &= \frac{\pi m}{2\sqrt{\alpha}}\, ,
\end{align} 
where $I_0$ is a modified Bessel function of the first kind.
The latter has asymptotic behavior
\begin{align}
I_0(y) \approx \frac{e^y}{\sqrt{2\pi y}}\left(1+\frac{1}{8y} + \cdots  \right)
\end{align}
for large $y$ within the domain $|\arg(y)|<\pi/2$.
Therefore, as expected for non-zero $\rho$, \eqref{vansh} vanishes as $\Lambda$ approaches $\Lambda_m$ along the real axis, 
\begin{align}
\left.\frac{\partial {\mathbb S}}{\partial \sigma}\right|_{\sigma=0} \sim \Lambda-\Lambda_m\, ,
\end{align}
whereas for $\rho=0$,
\begin{align}\label{polemo}
\left.\frac{\partial {\mathbb S}}{\partial \sigma}\right|_{\sigma=0} \sim \frac{1}{(\Lambda-\Lambda_m)^2}
\end{align}

Despite the lack of a divergence of $\langle\Omega\rangle$ as $\Lambda\rightarrow \Lambda_m$ along the real axis,  
the perturbation introduces an essential singularity of ${\mathbb S}$ at $\Lambda=\Lambda_m$.  This is not immediately obvious, since the essential singularities of $e^{-y}$ and $I_0(y)\sim e^{y}$ at $y=\infty$ apparently cancel in the product $e^{-y}(I_0(y) - I'_0(y))$ which appears in \eqref{vansh} with $y\sim 1/(\Lambda-\Lambda_m)^2$.  
The signature of this product having an essential singularity is that although $f(y)\equiv e^{-y}I_0(y)$ vanishes as $y\rightarrow\infty$ along the positive real $y$ axis,  it diverges exponentially approaching $\infty$ along the negative real $y$ axis. To see this, note that $I_0(y)$ is even, so that \begin{align}
f(-y) = e^{y}I_0(y) = e^{2y}f(y) \sim \frac{e^{2y}}{\sqrt{2\pi y}} 
\end{align}
for large positive $y$. 
As noted above, for vanishing $\rho$ equation \eqref{polemo} generates motion of the ghost pole under variation of $\sigma$. 
Based on the appearance of the essential singularity in \eqref{vansh} for non-vanishing $\rho$,  together with the immovability of poles under local perturbations, we  propose that the pole is replaced with an essential singularity, moving as $\sigma$ is varied.

\begin{figure}[!h]
\center{
\includegraphics[width= 400pt]{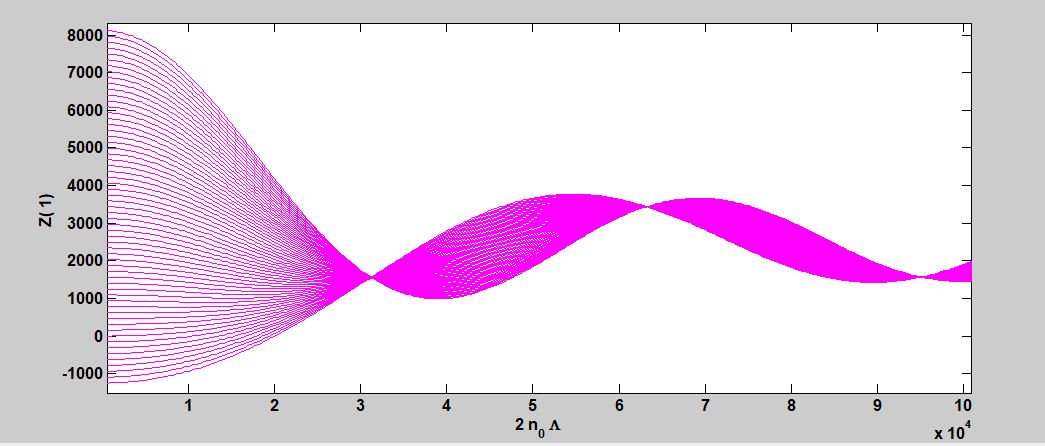}
\caption{Plot of endpoints $z(\tau=1)$ as a function of $2n_0\Lambda$ for a fixed $z(\tau=0)=3437.5$ and a range of initial `velocities' $\dot z(\tau=0)$, determined by numerical solution of the Euler Lagrange equations for the Lagrangian \eqref{Lagr} with $n(\vec X)^2 = n_0^2 - \alpha (Z-2500)^2 - \sigma (Z-2500)^2 \exp(-\rho (Z-2500)^2),\,\, n_0 = 6.7114*10^{-4},\,\, \alpha = 4.6124*10^{-15},\,\, \sigma = 5*10^{-4}, \rho = 2.56*10^{-6}$.  The parameter values are chosen so that the characteristic scales are similiar to those in ocean acoustics in units of meters and seconds. The Dirichlet problem degenerates at certain $\Lambda$, meaning that there are continuous classes of solutions with the same endpoints.  However the apparent collapse of the endpoints at special values of $\Lambda$ is not total, as can be seen zooming in as in figure \ref{fig:ButterflyLambdaZoom}}
\label{fig:ButterflyLambda}
}
\end{figure}

\begin{figure}[!h]
\center{
\includegraphics[width= 300pt]{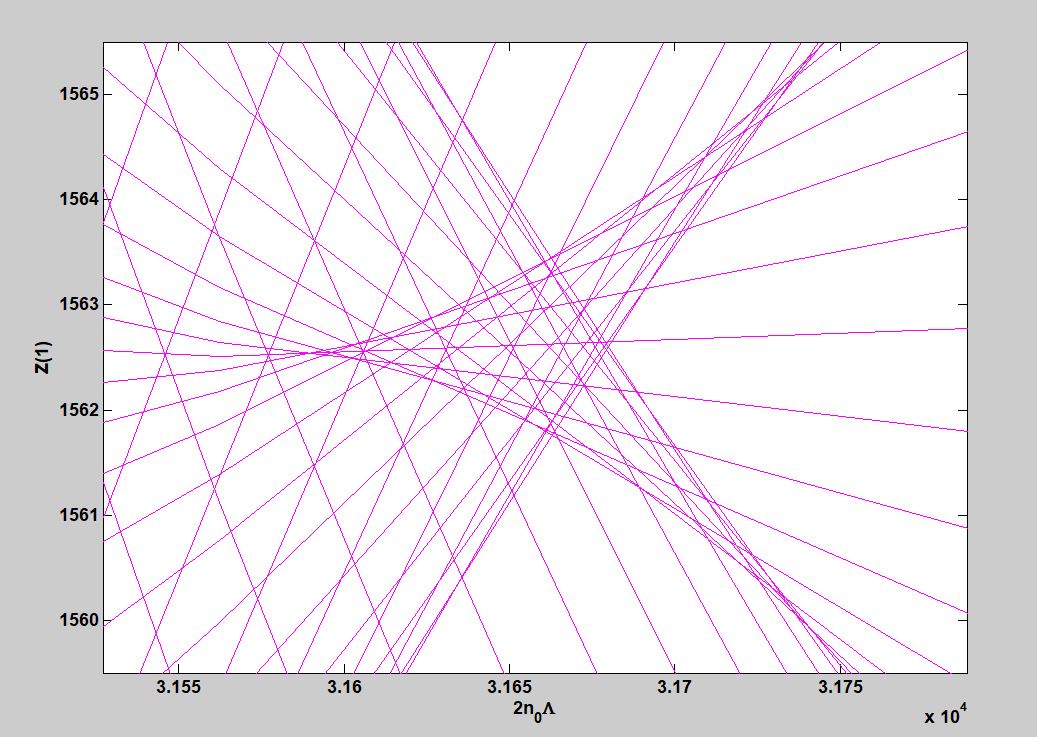}
\caption{A zoomed in view of figure \ref{fig:ButterflyLambda}, focusing on the neighborhood of a singular point. The pattern is that of an $A_5$ or `butterfly' catastrophe.  This pattern is nearly replicated in the fan of rays shown in figure \ref{fig:ButterflyRayZoom}.}  
\label{fig:ButterflyLambdaZoom}
}
\end{figure}

\begin{figure}[!h]
\center{
\includegraphics[width= 400pt]{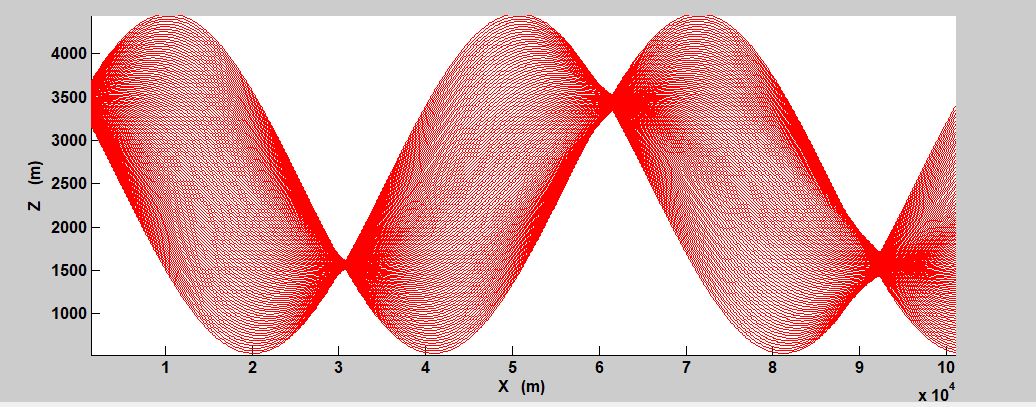}
\caption{Fan of rays for the same initial $z$ and parameter values of figure \ref{fig:ButterflyLambda}, and a range of launch angles.}
\label{fig:ButterflyRay}
}
\end{figure}

\begin{figure}[!h]
\center{
\includegraphics[width= 300pt]{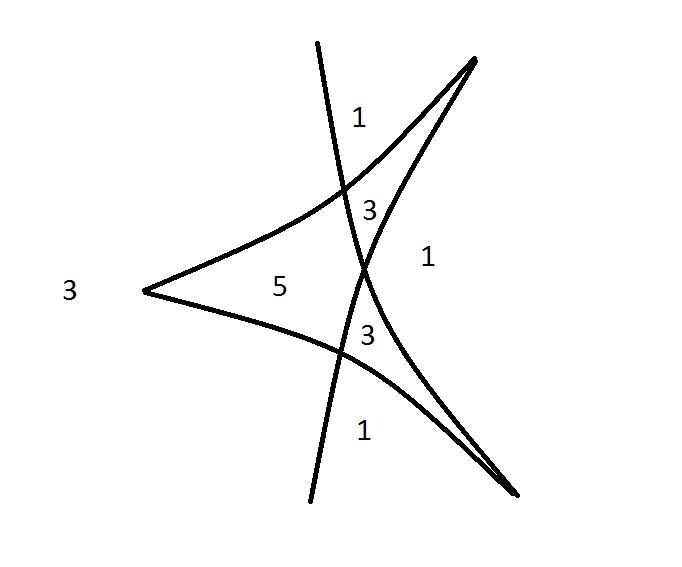}
\caption{Illustration of a two dimensional section of the four dimensional butterfly caustic.  In the present context, the two dimensions of the section are spatial, whereas the transverse dimensions are related  to  parameters of the problem.  The number of real rays within each domain is indicated. }
\label{fig:ButterflySketch}
}
\end{figure}

In the previous section, it was argued that the Dirichlet problem for the Euler Lagrange equations derived from \eqref{Lagr} collapses at ghost poles, at which finite solutions only exist at the spatial location of vanishing residue, or ghost sources.  Analogous behavior can be investigated numerically for the index of refraction \eqref{sigdef}, for which we have proposed that poles are replaced by essential singularities at non-zero $\rho$ and non-zero $\sigma$.  For arbitary initial `velocities' $\dot z(\tau=0)$ and fixed $z(\tau=0)$, one finds that continuous subsets of initial $\dot z$ yield the same $z(\tau=1)$ at particular values of $\Lambda$.  This differs from the exactly soluble cases in which the Dirichlet problem collapses completely at $\Lambda$ corresponding to poles. For a generic endpoint $z(\tau=1)$, there is now a class of solutions which diverges and another class which remains finite as $\Lambda$ approaches essential singularities of ${\mathbb S}(\Lambda)$.  Figure \ref{fig:ButterflyLambda} shows the behavior of endpoints $z(\tau=1)$ as a function of $2n_0\Lambda$ for a range of initial $\dot z(\tau=0)$. 
Looking at this figure, there are clearly values of $\Lambda$ at which the endpoints collapse, but not completely as can be seen zooming in, as shown in figure \ref{fig:ButterflyLambdaZoom}.  Remarkably, the pattern in the $\{\Lambda,z(\tau =1)\}$ space is an $A_5$ or butterfly catastrophe!  Considering the ray fan, shown in figure \ref{fig:ButterflyRay},  there is a butterfly caustic which very closely resembles that in the $\{\Lambda,z(\tau=1)\}$ space, which can be seen after zooming in figure \ref{fig:ButterflyLambdaZoom}. Since there are two spatial dimensions and the butterfly caustic is codimension 4, the pattern in the figure is a slice of the caustic, with the two un-displayed dimensions corresponding to parameters other than spatial directions.  A sketch of the caustic surface is shown in figure \ref{fig:ButterflySketch}, indicating the number of real rays within each domain.

\begin{figure}[!h]
\center{
\includegraphics[width= 300pt]{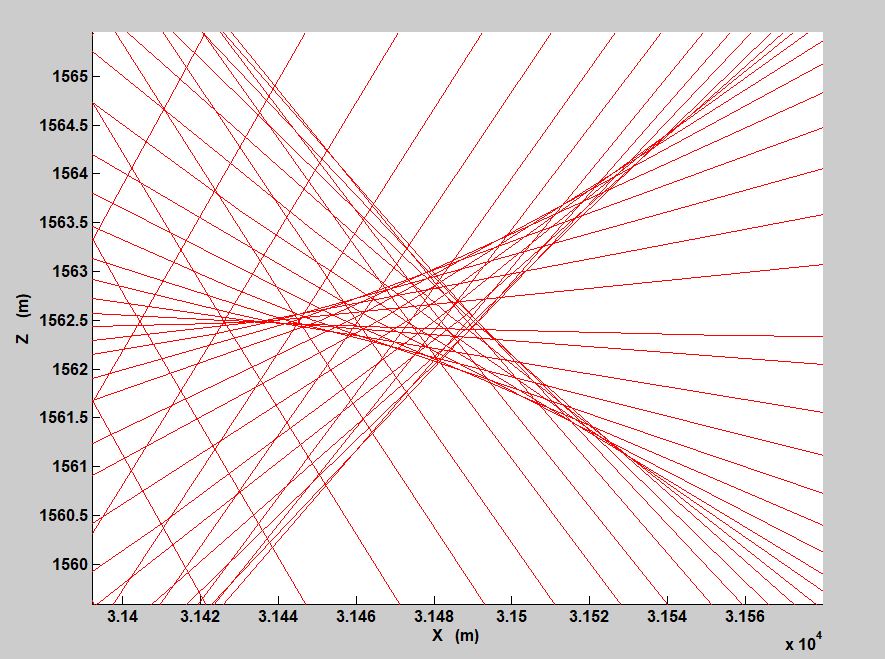}
\caption{Zooming into a an intense caustic region of figure \ref{fig:ButterflyRay} shows a butterfly caustic, containing 3 cusps and 3 self-intersections.  The pattern closely mirrors figure \ref{fig:ButterflyLambda}, reflecting a map between higher order caustics and singularities of the Dirichlet problem for the Lagrangian \eqref{Lagr}, which in turn corresponds to singularities of the einbein action. }
\label{fig:ButterflyRayZoom}
}
\end{figure}

We repeat our emphasis that the fan of rays and the fan of curves corresponding to endpoints $Z(1)$ as a function of $\Lambda$ are in principle in-equivalent, living in very different spaces, and that the Euler Lagrange equations for \eqref{Lagr} are not ray equations.  The resemblance between figures \ref{fig:ButterflyLambda} and \ref{fig:ButterflyRay} applies only to the neighborhood of the singular points. The degenerations of the Dirichlet problem in the einbein description is clearly intimately related to caustics in the ray description.  The relation between singularities of the einbein action and caustics was first proposed in \cite{GLefschetz}, but was limited to simple poles and the $A_3$ catastrophe or cusp caustic.  The numerical and perturbative analysis here suggests that this relation generalizes to include essential singularities of the einbein action and higher order $A_{n>3}$ caustics. 

We speculate that the correspondence between singular points in the Dirichlet problem for the Lagrangian \eqref{Lagr} and the $A_{n>2}$ caustics can be expressed more generally in terms of the local coincidence of two surfaces.  One of these is the surface spanned by the ray fan $\vec X_{ray}(\vec P_0,s)$ where $P_0$ are the intial canonical momenta (or velocities) and $s$ parameterizes the ray paths.  The other surface is given by $\vec{\mathbb X}(\vec P_0,\Lambda)$, where $\vec{\mathbb X}$ is the endpoint of the path $\vec X(\tau=1)$ where $\vec X(\tau)$ satisfies the Euler Lagrange equations of \eqref{Lagr}. In the latter case, there is no Eikonal constraint $\vec P^2 - n(\vec X_0)^2=0$, but it can be imposed on the initial $\vec P_0$ so that the two surfaces have the same dimension.    
For the ray fan, one can choose the parameter $s$ to be be the physical travel time, which is proportional to the real part of the  einbein action at a critical point  \cite{GLefschetz}, at which $d{\mathbb S}/d\Lambda =0$.  Alternatively, one can choose $s$ to be the corresponding critical value $\Lambda_{crit}$. One then has the two surfaces $\Sigma_{1,2}$,
\begin{align}\label{surf2}
\Sigma_1:& \,\,\, \vec X_{ray}(\vec P_0,\Lambda_{crit}) \nonumber\\
\Sigma_2:& \,\,\, \vec {\mathbb X}(\vec P_0, \Lambda)
\end{align}
We propose that $\Sigma_1$ and $\Sigma_2$ are coincident at $A_{n>2}$ caustics.
Approaching an $A_{n>2}$ caustic, $\Lambda_{crit}\rightarrow \Lambda_p$, where $\Lambda_p$ is a pole or other singularity. 
Note that these surfaces are not globally equivalent, even though the Eikonal condition has been imposed when generating $\Sigma_2$, since $\mathbb X$ corresponding to an endpoint evaluated at fixed $\tau=1$. The equivalence of the surfaces is claimed only in the neighborhood of $A_{n>2}$ caustics.

Recall that, for the exactly soluble model discussed in \ref{polepairs}, as well as the Munk profile, the position of cusps was given by 
\begin{align}
z&=z_{ghost} \label{loclz}\\ 
x&\approx 2n_{eff}\Lambda_p \label{loclx}\, .
\end{align}
The critical points of the einbein action \eqref{chanac} are more easily computed along a ghost source, at which the associated pole terms in the action can be neglected\footnote{The counting of critical points along a ghost source but away from a caustic is subtle, in that one critical point survives while two critical points coallesce and annihilate, without yielding a caustic or changing the number of rays.  The logarithmic branch points of $\mathbb S(\Lambda)$ are of critical importance in this case.  These subtleties are  discussed in detail in \cite{GLefschetz} }. For $z$ along the locus of a ghost source, the action at the critical point is ${\mathbb S} \approx k_0 2 n(z_{ghost})^2 \Lambda_{crit}$. Translating this to a distance,
\begin{align}
D\equiv \frac{{\mathbb S}}{k(z_{ghost})} = 2n(z_{ghost})\Lambda_{crit}\,
\end{align}
consistent with \eqref{loclx} upon approaching the caustic, $\Lambda_{crit}\rightarrow\Lambda_p$.  The coincidence of the curves $\Sigma_1$ and $\Sigma_2$ at caustics may be though of a generalization of the caustic localization \eqref{loclz} and \eqref{loclx}.  In fact the latter localization is suitable for an index of refraction characteristic of problems in ocean acoustics, for which the long range propagation is mostly horizontal (in the $x$ direction) and $\Lambda_p$ is much larger than the characteristic vertical scales.

As argued above, there ought to be large classes of local perturbations which evade the diverging trajectories of solutions of the Euler Lagrange equations for \eqref{Lagr} as $\Lambda$ approaches a pole.  In such cases one does not expect the pole or its residue to vary.  Cusps may nevertheless slide along the fixed locus of vanishing residue, or ghost sources.  Indeed this is seen numerically for a variety of perturbations. 
In such cases it is legitmate to consider 
the effect of perturbations on a Laurent series expansion of the einbein action about a pole. Such an approach can not capture essential singularities of the einbein action, which vanish to all orders in the Laurent series.  In the Laurent series solution, the equations governing the residues of a pole are independent of the index of refraction, as will demonstrated below. However higher order terms are sensitive to perturbations of the index of refraction. 

The Laurent series solution was described in \cite{GLefschetz}, and we repeat it here.
One starts by assuming a solution to the Schr\"oedinger equation \eqref{Schrod} of the form 
\begin{align}
\Psi = &\left(\frac{k_0}{4\pi i \tilde\Lambda}\right)^{D/2}e^{i\bar S} \nonumber \\
&\bar S = 
ik_0( \gamma_{-1}\frac{1}{\tilde\Lambda} + \gamma_0+ \gamma_1 \tilde\Lambda + \cdots ) \, ,
\end{align}
where $\tilde\Lambda\equiv \Lambda-\mu$ for some $\mu$, and $D$ is an integer. The Schr\"oedinger equation becomes
\begin{align}
ik_0 \frac{\partial}{\partial\tilde\Lambda}\Psi(\Lambda)+ \left(\vec\nabla_x^2 + k_0^2n^2\right)\Psi(\Lambda) = \left(\Upsilon_{-2} \frac{1}{\tilde\Lambda^2} + \Upsilon_{-1} \frac{1}{\tilde\Lambda} + \cdots \right)\Psi\, , 
\end{align}
where one insists $\Upsilon_I=0$ for all $I$. The equations derived rom $\Upsilon_{-2}=0$ is
\begin{align}\label{nabsqd}
-(\vec \nabla_x \gamma_{-1})^2 + \gamma_{-1}=0\, 
\end{align}
while $\Upsilon_{-1}=0$ implies
\begin{align}
\vec\nabla^2\gamma_{-1}&-\frac{D}{2} =0 \label{eqone}\\
\vec\nabla\gamma_0\cdot&\vec\nabla\gamma_{-1}=0\, . \label{eqtwo}
\end{align}
The index of refraction does not appear at all at this order.  
The solution of \eqref{nabsqd} and \eqref{eqone} is 
\begin{align}
\gamma_{-1} = \frac{1}{4}R^2
\end{align}
where $R$ is the distance to some, possibly curved, codimension $D$ surface,
\begin{align}
R^2= \sum_{j=1}^D \xi_j^2\, ,
\end{align} 
and the metric can be written as
\begin{align} \label{foliate}
d\vec x^2 = g_{ij}(\vec\xi,\vec\sigma)d\sigma^id\sigma^j+ \sum_{j=1}^D \xi_j^2 + \, ,
\end{align}
for coordinates $\vec\sigma$ parameterizing the surface.  This surface corresponds to the a source or ghost source. 
Higher orders show dependence on $n(\vec X)$.  For instance, vanishing $\Upsilon_0$ requires
\begin{align}
\gamma_1 + 2\vec\nabla\gamma_{-1}\cdot\vec\nabla\gamma_1 + (\vec\nabla\gamma_0)^2 - n^2- \frac{i}{k_0}\vec\nabla^2\gamma_0 =0\, .
\end{align} 
Therefore $\gamma_1$ is given by
\begin{align} \label{gamone}
\gamma_1 = \left( 1 + 2(\vec\nabla\gamma_{-1})\cdot\vec \nabla\right)^{-1}\left(n^2 - (\vec\nabla\gamma_0)^2 + \frac{i}{k_0}\vec\nabla^2\gamma_0\right)
\end{align}
or 
\begin{align}\label{solg}
\gamma_1(\vec \sigma,\vec \xi) = \frac{1}{|{\vec\xi}|} \int ds \left(n^2 - (\vec\nabla\gamma_0)^2 + \frac{i}{k_0}\vec\nabla^2\gamma_0\right)
\end{align}
where the integral is along the line segment from $(\vec \sigma,\vec\xi=0)$ to $(\vec \sigma,\vec\xi'=\vec\xi)$.  The effect of perturbations of $n$ is then 
\begin{align}
\gamma_1 \rightarrow \gamma_1+ \epsilon \frac{1}{R} \int d\xi' \Omega \, .
\end{align}
Note that $\gamma_1$ is the coefficient of the term in the einbein action which is linear in $\Lambda$, much like $n_{eff}$ in \eqref{effein}, to which it is presumably closely related. Similiar expressions apply to the perturbations of higher order terms in the Laurent series.  Perturbative effects on this term involve non-local averages of the variation in the index of refraction, which may therefore become negligible for  local perturbations at sufficiently large $R$.  
Assuming the location of cusps along ghost sources is proportional to the product of a the ghost pole $\Lambda_p$ with a non-local average of the index of refraction $n_{eff}$, then this position will also be relatively robust at distances which are large compared to the scale of perturbations.

\section{Conclusions}

We have interpreted the non-smooth $A_{n>2}$ caustics of the Helmholtz equation in terms of singularities of the einbein action.  The
detailed mathematics of many of the claimed relations  remains to be worked out.  The correspondence between the $A_{n>2}$ caustics, poles and essential singularities of the einbein action, and degenerations of the Dirichlet problem for the Lagrangian \eqref{Lagr} are largely conjectural, based on a combination of numerical observation, exactly soluble cases and perturbative arguments.  If the arguments hold up to a more rigorous study, there are remarkable implications for the effect of variations of the index of refraction on the non-smooth caustics, based on the relation between deformations of the singularities and the associated aspects of caustic geometry.

\section{Acknowledgements}

This work was supported by DARPA.\\
Distribution statement ``A'' (Approved for Public Release, Distribution Unlimited)
\vskip10pt
\noindent {\it The views, opinion, and/or findings expressed are those of the authors and should not be interpreted as representing official views or policies of the Department of Defense or the U.S. Government.}

\end{document}